# All-optical magnetometric characterization of the antiferromagnetic exchange-spring system Mn$_2$Au|Py by terahertz spin-torques

Y. Behovits[1], A. L. Chekhov[1], B. Rosinus Serrano[1], A. Ruge[1], S. Reimers[2], Y. Lytvynenko[2,3], M. Kläui[2], M. Jourdan[2] and T. Kampfrath[1]

1. Department of Physics, Freie Universität Berlin, 14195 Berlin, Germany
2. Institute of Physics, Johannes Gutenberg-Universität Mainz, 55099 Mainz, Germany
3. Institute of Magnetism of the NAS and MES of Ukraine, 03142 Kyiv, Ukraine

**Abstract**

Antiferromagnetic materials have great potential for spintronic applications at terahertz (THz) frequencies. However, in contrast to ferromagnets, experimental studies of antiferromagnets are often challenging due to a lack of straightforward external control of the Néel vector $\boldsymbol{L}$. Here, we study an AFM|FM stack consisting of an antiferromagnetic metal layer (AFM) of the novel material Mn$_2$Au and a ferromagnetic metal layer (FM) of NiFe. In this exchange-spring system, $\boldsymbol{L}$ of AFM Mn$_2$Au can be controlled by the application of an external magnetic field $\boldsymbol{B}_{\mathrm{ext}}$. To characterize the AFM|FM stack as a function of the quasi-static $\boldsymbol{B}_{\mathrm{ext}}$, we perform THz-pump magneto-optic probe experiments. We identify signal components that can consistently be explained by the in-plane antiferromagnetic magnon mode excited by field-like Néel spin-orbit torques (NSOTs). Remarkably, we find that the $\boldsymbol{B}_{\mathrm{ext}}$- and THz-pump-induced changes in the optical response of the sample are dominated exclusively by the spin degrees of freedom of AFM. We fully calibrate the magnetic circular and magnetic linear optical birefringence of AFM and extract the efficiency of the NSOTs. Finally, by selective excitation of domains with different orientation of $\boldsymbol{L}$, we are able to determine the relative volume fraction of 0°, 90°, 180° and 270° domains distribution during the quasi-static reversal of $\boldsymbol{L}$ by $\boldsymbol{B}_{\mathrm{ext}}$. Our insights are an important prerequisite for future studies of ultrafast coherent switching of spins by THz NSOTs and show that THz-pump magneto-optic-probe experiments are a powerful tool to characterize magnetic properties of antiferromagnets.

**Figures**

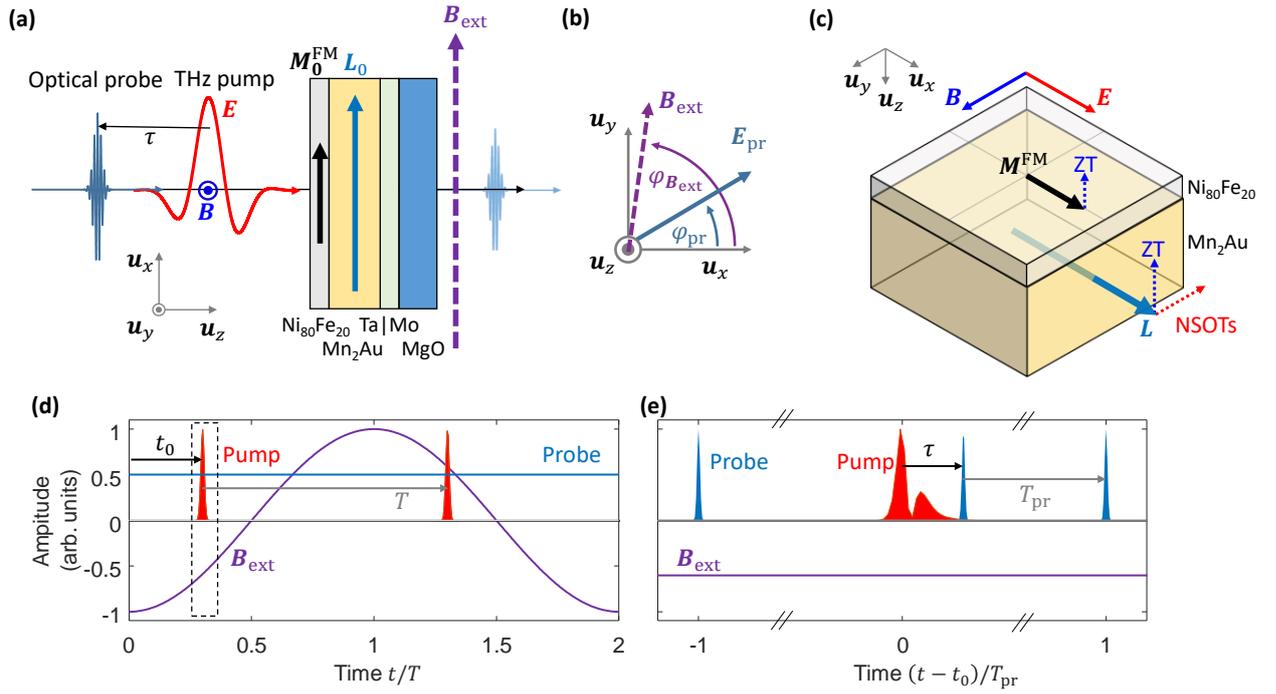

**Fig. 1: Schematic of experiment and torque-driven dynamics. (a)** Schematic of the experiment. A THz pump pulse is incident on the sample, and the resulting spin dynamics is interrogated by an optical probe through magneto-optic effects. The sample is a thin-film stack sub|seed|AFM|FM, where the substrate slab (sub) is MgO(500 µm), the seed layer is Mo(20 nm)|Ta(13 nm), the antiferromagnetic layer AFM is Mn₂Au(50 nm), and the ferromagnetic layer FM is Ni₈₀Fe₂₀(10 nm). **(b)** Probe polarization angle $\varphi_{\mathrm{pr}} = \sphericalangle(\boldsymbol{E}_{\mathrm{pr}}, \boldsymbol{u}_x)$, where $\boldsymbol{E}_{\mathrm{pr}}$ is the probe electric field and $\varphi_{\boldsymbol{B}_{\mathrm{ext}}}$ the angle of the external magnetic field $\boldsymbol{B}_{\mathrm{ext}}$. **(c)** Possible Zeeman torque (ZT) and Néel spin-orbit torques (NSOTs) on layers FM and AFM with order parameters $\boldsymbol{M}^{\mathrm{FM}}$ and $\boldsymbol{M}$, $\boldsymbol{L}$, respectively. **(d)** Timing diagram of one cycle $[0, 2T[$ of the stroboscopic measurement. The purple line shows the external magnetic field $\boldsymbol{B}_{\mathrm{ext}}(t)$ with a period of $2T$, where $T = 1\,\mathrm{ms}$. The THz pump pulses (red; duration exaggerated) arrive with repetition period $T$ at times $t_0$ and $t_0 + T$ relative to the magnetic-field minimum $-B_{\mathrm{ext}0}$. The pump pulse width is exaggerated for visibility. The probe pulses are spaced equidistantly by the repetition period $T_{\mathrm{pr}} = 12.5\,\mathrm{ns} \ll T$ and arrive quasi-continuously. **(e)** Magnified view of the dashed-line box in panel (d). It shows the THz pump-pulse intensity $E^2(t)$ (red) and a probe pulse (blue) that arrives with a delay $\tau$ after the pump. One preceding and subsequent probe pulses at times $t_0 + \tau \pm T_{\mathrm{pr}}$ are shown, too. On the time scale $\tau \ll T$ of the pump-probe experiment, $\boldsymbol{B}_{\mathrm{ext}}$ is approximately constant.

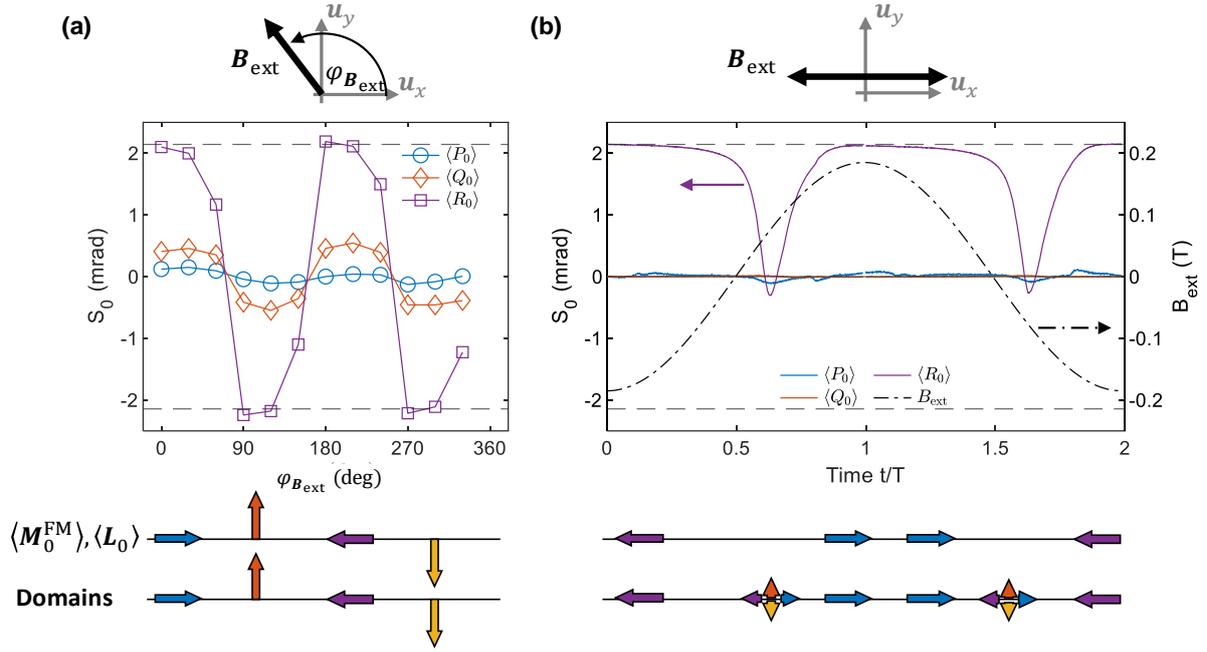

**Fig. 2: Magneto-optic signal vs quasi-static external magnetic field. (a)** Probe-polarization rotation signals $S_0$ vs azimuthal angle $\varphi_{B_{\text{ext}}}$ of the static in-plane magnetic field $\boldsymbol{B}_{\text{ext}} = -B_{\text{ext0}}\left(\cos\varphi_{B_{\text{ext}}}\boldsymbol{u}_x + \sin\varphi_{B_{\text{ext}}}\boldsymbol{u}_y\right)$. Signal components $\langle P_0 \rangle, \langle Q_0 \rangle, \langle R_0 \rangle$ are extracted according to Eq. (4) from 4 linear polarization angles (Appendix A, Figure S 2). The mean signal is set to zero. The first schematic below shows the orientation of the magnetization $\langle \boldsymbol{M}_0^{\text{FM}} \rangle$ of layer FM (Py) and the Néel vector $\langle \boldsymbol{L}_0 \rangle$ of layer AFM (Mn$_2$Au) at $\varphi_{B_{\text{ext}}} = 0, 90, 180, 270°$. The second schematic shows the statistics of the local $\boldsymbol{L}_0$ assuming saturation. **(b)** Signal $S_0(t)$ [Eq. (4)] versus time $t$ in an external field with quasi-statically time-varying amplitude $B_{\text{ext}}(t)$ along $\boldsymbol{u}_x$, i.e., $\boldsymbol{B}_{\text{ext}} = B_{\text{ext}}(t)\boldsymbol{u}_x$ (black line). Components $\langle P_0 \rangle, \langle Q_0 \rangle, \langle R_0 \rangle$ are extracted according to Eq. (4) from 8 linear polarization angles (Figure S 12). For $\langle P_0 \rangle$ and $\langle Q_0 \rangle$, the signal is set to zero for $t = 0$. For signal $\langle R_0 \rangle$, the amplitude at $t = 0$ is set to the maximum value of the corresponding signal in panel (a) (dashed lines). This procedure assumes that $\langle \boldsymbol{M}_0^{\text{FM}} \rangle$ and $\langle \boldsymbol{L}_0 \rangle$ are saturated by $\boldsymbol{B}_{\text{ext}}$. The first schematic below shows the extracted $\langle \boldsymbol{M}_0^{\text{FM}} \rangle$ and $\langle \boldsymbol{L}_0 \rangle$ at various points in time. The second one displays the expected distribution of the local $\boldsymbol{L}_0$ over 4 domains. At $t/T \approx 0.6$, a multi-domain state is expected, where $\langle \boldsymbol{M}_0^{\text{FM}} \rangle$ and $\langle \boldsymbol{L}_0 \rangle$ vanish and the domains are equally populated, as indicated by the four arrows.

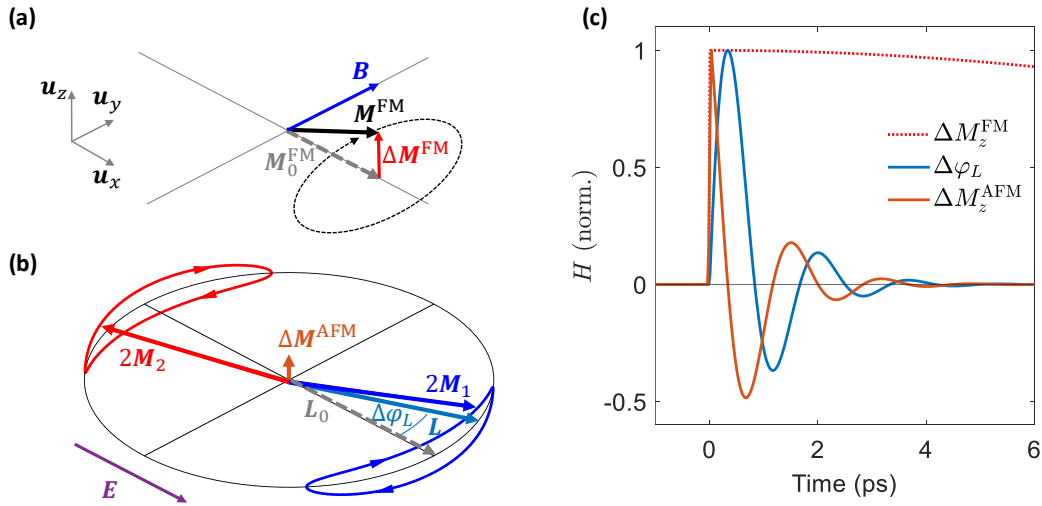

**Fig. 3**: **Expected spin dynamics and response functions linear in driving THz field $F = (E, B)$.**
**(a)** Spin dynamics in the ferromagnetic Permalloy (Py) layer driven by Zeeman torques (ZT) from the THz magnetic field $B$. The magnetization $M_0^{\mathrm{FM}}$ in the absence of a pump pulse is given by gray dashed arrow. ZT leads to out-of-plane ($\parallel u_z$) dynamic $\Delta M^{\mathrm{FM}}$ and a subsequent precession of $M^{\mathrm{FM}} \approx M_0^{\mathrm{FM}} + \Delta M^{\mathrm{FM}}$. **(b)** Spin dynamics in the antiferromagnetic layer driven by Néel spin-orbit torques (NSOTs) from the THz electric field $E$. NSOTs lead to an out-of-plane $\Delta M^{\mathrm{AFM}}$ and in-plane deflection of the Néel vector by an angle $\Delta\varphi_L$. **(c)** Resulting impulse-response functions $H$. In FM, magnetization $\Delta M_z^{\mathrm{FM}}$ is induced by Zeeman torque (red-dashed line). In AFM, NOTs cause an in-plane deflection $\Delta\varphi_L$ of the Néel vector (blue line) and an out-of plane magnetization $\Delta M_z^{\mathrm{AFM}}$ (orange line).

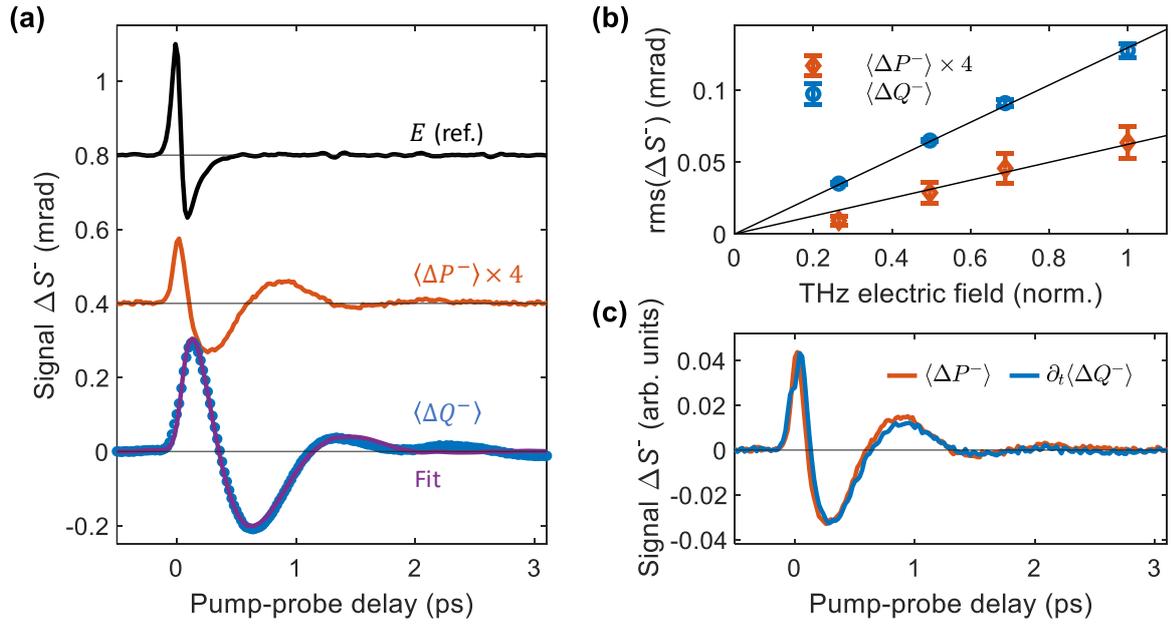

**Fig. 4: Pump-probe signal components odd in both THz pump field $F$ and external magnetic field $B_{\text{ext}}$.** (a) Ellipticity-signal components $\langle \Delta P^- \rangle$ (orange line) and $\langle \Delta Q^- \rangle$ (blue circles), along with the THz pump electric field (black). Signals are extracted according to Eq. (4) from 8 linear polarization angles (Appendix A). Raw data are shown in Figure S 3. Signals are scaled and vertically offset for clarity. Purple line shows a fit to signal $\langle \Delta Q^- \rangle$ according to the response function in Table 2. (b) RMS of the signals (Figure S 10) vs pump electric-field strength. The black lines are linear fits. (c) Signal $\langle \Delta P^- \rangle$ (orange line) and scaled numerical derivative signal $\partial_t \langle \Delta Q^- \rangle$ (blue line).

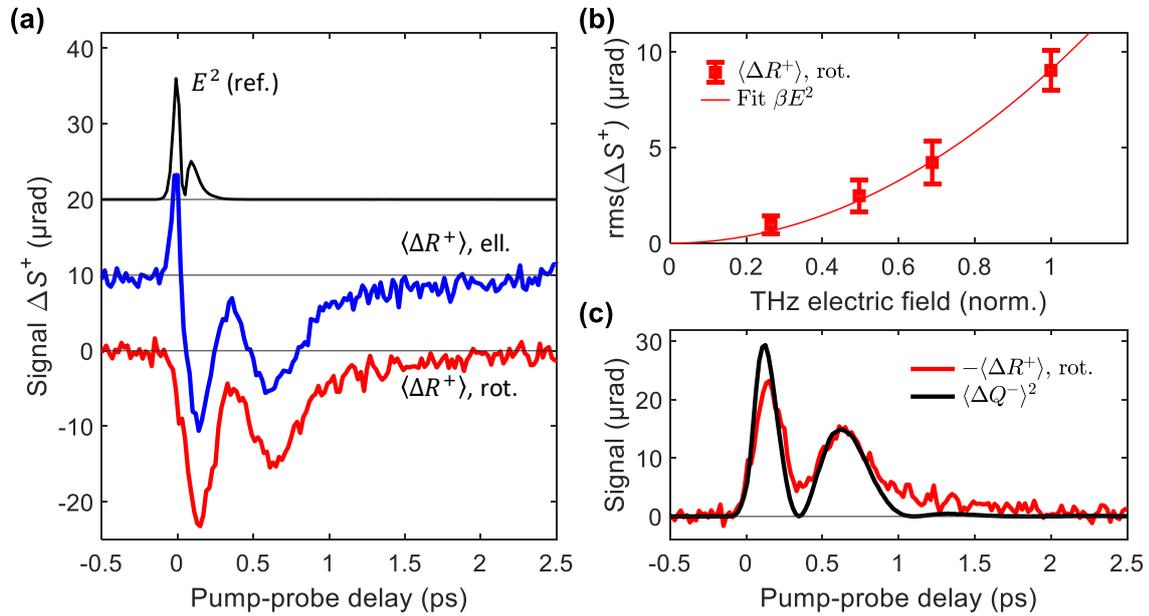

**Fig. 5: Pump-probe signal components even in both THz pump field $F$ and external magnetic field $B_{\text{ext}}$.** (a) Pump-probe signal component $\langle \Delta R^+ \rangle$ even in both THz pump field $F$ and external magnetic field $B_{\text{ext}}$. The signal was extracted according to Eq. (4) from 8 linear polarization angles (Figure S 3, Figure S 4) and measured as probe ellipticity (blue line) and rotation (red line). The black solid line shows the squared pump electric field $E^2(\tau)$. (b) RMS of the rotation signal (Figure S 11) vs pump electric-field strength and quadratic fit. (c) Signal $-\langle \Delta R^+ \rangle$ (rotation, red line) and scaled signal $\langle \Delta Q^- \rangle^2$ (black line).

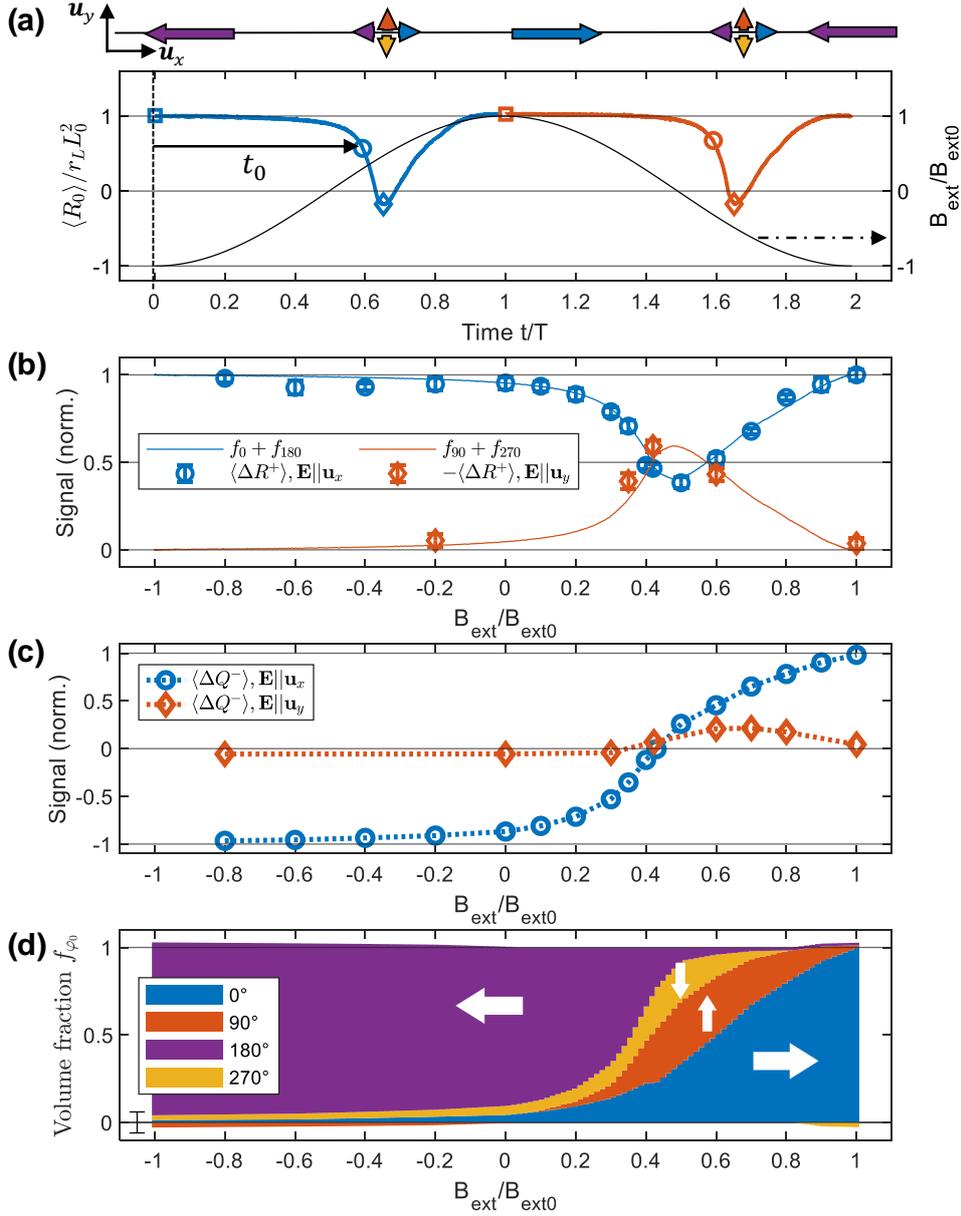

**Fig. 6: Probing domain statistics. (a)** Concept of the experiment. The schematic shows the expected qualitative domain distribution when the amplitude $B_{\text{ext}}(t)$ of the external magnetic field $\boldsymbol{B}_{\text{ext}} = B_{\text{ext}}\boldsymbol{u}_x$ increases quasi-statically from $-B_{\text{ext0}}$ at $t = 0$ to $+B_{\text{ext0}}$ at $t = T = 1\,\text{ms}$ and subsequently returns to $-B_{\text{ext0}}$ at $t = 2T$ (solid black line). The graph shows $\langle\cos(2\varphi_{L_0})\rangle = \langle R_0\rangle/r_L L_0^2$ for one cycle of increasing (blue line) and decreasing (orange line) $B_{\text{ext}}$, analogous to Fig.2b. To characterize the domain volume fractions $f_{0°}, f_{90°}, f_{180°}, f_{270°}$ at a given time $t_0$, we separately apply THz NSOTs to domains with $\boldsymbol{L}_0 \parallel \boldsymbol{u}_x$ and $\boldsymbol{L}_0 \parallel \boldsymbol{u}_y$ and probe magneto-optically. The THz pump pulses arrive at times $t_{\text{pu}} = t_0$ and $t_{\text{pu}} = t_0 + T$. Symbols indicate $t_{\text{pu}}$ instances, e.g., $t_{\text{pu}} = 0$ (blue square) and $t_{\text{pu}} = T$ (red square), where $B_{\text{ext}}$ has maximum magnitude, i.e., $B_{\text{ext}} = \pm B_{\text{ext0}}$. **(b)** Pump-probe signal amplitude $\langle\Delta R^+\rangle$ even in $B_{\text{ext}}$ and, thus, driving THz field vs amplitude $B_{\text{ext}}$ for electric field $\boldsymbol{E} \parallel \boldsymbol{u}_x$ (blue circles) and $\boldsymbol{E} \parallel \boldsymbol{u}_y$ (orange diamonds) [Figure S 11]. For comparison, the volume fractions $f_{0°} + f_{180°}$ (blue line) and $f_{90°} + f_{270°}$ (orange line) as derived from quasi-static signals are shown [Eq. (20)]. **(c)** Amplitude of $\langle\Delta Q^-\rangle$ for electric field $\boldsymbol{E} \parallel \boldsymbol{u}_x$ (blue circles) and $\boldsymbol{E} \parallel \boldsymbol{u}_y$ (orange diamonds) [see Figure S 10]. **(d)** Extracted volume fractions $f_{0°}, f_{90°}, f_{180°}$ and $f_{270°}$ of domains in the 4 easy axis directions vs external magnetic field $B_{\text{ext}}$. The $f_{\varphi_{L_0}}$ are obtained by combining the amplitude of quasi-static signals in panel (a) and the pump probe signals in panel (c) according to Appendix E. White arrows indicate the direction of the Néel vector $\boldsymbol{L}_0$. The error bar shows the uncertainty resulting from the small offset in the orange curve of panel (c).

## Tables

### Table 1:

**Table of symmetry-allowed signal components [Eq. (5)].** $L$ is only present antiferromagnetic layer, while $M$ is generally present in the ferromagnetic and the antiferromagnetic layer. The coefficients $p(r)$, $q_L(r)$, $q_M(r)$, $r_L(r)$, $r_M(r)$ quantify the sensitivity to the local $L(r)$ and $M(r)$ at position $r$ and scale with the sample thickness. The total signal $\langle X \rangle$ arises from averaging of $X = P, Q, R$ over the probed volume $V$ according to $\langle X \rangle = (1/V) \int_V d^3r\, X(r)$.

| Component | Type | Related $\varphi_{\mathrm{pr}}$ dependence | Contribution due to changes exclusively in spin degrees of freedom $\mathcal{S}$ | Non-spin contribution |
|---|---|---|---|---|
| $P$ | MCB | 1 | $P_\mathcal{S} = p_M M_z$ | $P_\mathcal{N} = P - P_\mathcal{S}$ |
| $Q$ | MLB | $\cos(2\varphi_{\mathrm{pr}})$ | $Q_\mathcal{S} = 2q_L L_x L_y + 2q_M M_x M_y$ $= q_L L_{\parallel 0}^2 \sin(2\varphi_L) + q_M M_{\parallel 0}^2 \sin(2\varphi_M)$ | $Q_\mathcal{N} = Q - Q_\mathcal{S}$ |
| $R$ | MLB | $\sin(2\varphi_{\mathrm{pr}})$ | $R_\mathcal{S} = r_L(L_x^2 - L_y^2) + r_M(M_x^2 - M_y^2)$ $= r_L L_{\parallel 0}^2 \cos(2\varphi_L) + r_M M_{\parallel 0}^2 \cos(2\varphi_M)$ | $R_\mathcal{N} = R - R_\mathcal{S}$ |

### Table 2:

**Possible spin dynamics** in linear response to a terahertz pump field $F = (E, B)$. Here, $\Theta(t)$ denotes the Heaviside step function, and $M_0, L_0 \parallel u_x$ is assumed.

| Magnetic order parameter $\Delta Y$ | Sensitive signal component | Driving field | Mechanism | Response function $H_{\Delta YX}$ is proportional to | Resonance frequency $\Omega/2\pi$ |
|---|---|---|---|---|---|
| $\Delta M_z^{\mathrm{FM}}$ | $\langle \Delta P \rangle$ | $B_y$ | ZT | $\cos(\Omega t)\, e^{-\Gamma t} \Theta(t)$ | $\sim 15$ GHz [1] |
| $\Delta L_z$ | - | $B_y$ | ZT | $H_{\Delta L_z B_y}(t) = (1/\Omega) \cos(\Omega t)\, e^{-\Gamma t} \Theta(t)$ | $\sim 3$ THz [2, 3] |
| $\Delta M_y^{\mathrm{AFM}}$ | $\langle \Delta R \rangle$ (small) | $B_y$ | ZT | $\partial_t H_{\Delta L_z B_y}$ | $\sim 3$ THz |
| $\Delta \varphi_L$ | $\langle \Delta Q \rangle, \langle \Delta R \rangle$ | $E_x$ | NSOTs | $H_{\Delta \varphi_L E_x}(t) = (1/\Omega) \sin(\Omega t)\, e^{-\Gamma t} \Theta(t)$ | $\sim 0.5$ THz [4] |
| $\Delta M_z^{\mathrm{AFM}}$ | $\langle \Delta P \rangle$ | $E_x$ | NSOTs | $\partial_t H_{\Delta \varphi_L E_x}$ | $\sim 0.5$ THz |

**Table 3:**

**Table of extracted magneto-optic coefficients.** All coefficients are given for the antiferromagnetic Mn$_2$Au layer with a thickness of 50 nm and for the probe wavelength of 800 nm. The MCB component is given for $M_z = L_0$.

| Component | Type | Coefficient | Rotation (mrad) | Ellipticity (mrad) |
|:---:|:---:|:---:|:---:|:---:|
| $P$ | MCB | $p_M L_0$ | 8 | 16 |
| $Q$ | MLB | $q_L L_0^2$ | 0.3 | 1.8 |
| $R$ | MLB | $r_L L_0^2$ | 2.2 | 2.2 |

**Main text**

**I Introduction**

Antiferromagnetic spin dynamics has attracted strongly increasing attention in the last years because of its potential to build spintronic devices that are robust against magnetic fields and operate at terahertz (THz) frequencies [5]. Importantly, the discovery of Néel spin-orbit torques (NSOTs) has opened up possibilities to manipulate antiferromagnetic order by electrical currents, in particular in the antiferromagnetic metals $Mn_2Au$ and CuMnAs. Remarkably, recent ultrafast THz-pump magneto-optic-probe experiments on $Mn_2Au$ thin films indicate that coherent 90° or even 180° rotation of the Néel vector $L$ through field-like NSOTs is possible, thereby paving the way to THz-rate information writing in antiferromagnets [4].

However, the observation of ultrafast switching of $L$ poses a number of unique challenges. (i) For experiments, a control parameter is needed that allows us to initialize $L$ in a well-defined state. Here, the robustness of antiferromagnets to external fields is, in fact, a disadvantage and prevents $L$ control in typical laboratory settings. (ii) Suitable ultrafast probes are required to detect the instantaneous magnetic order parameters $L$ and a possible transient magnetization $M$. In particular, we need to be able to differentiate opposite Néel-vector directions, i.e., $\pm L$. Protocols to calibrate signals relative to $L$ are highly desired. (iii) The torque on $L$ and, thus, the NSOT-driving THz pump electric fields need to be increased to overcome the switching barrier.

In this work, we use THz-pump magneto-optic probe experiments (Fig. 1a) to study an exchange-spring system, i.e., a FM|AFM stack, where the ferromagnetic layer FM of $Ni_{80}Fe_{20}$ (permalloy Py) and the antiferromagnetic layer AFM of $Mn_2Au$ are exchange-coupled [6]. Importantly, the FM magnetization can be used to control the $L$ landscape of AFM, thus overcoming challenge (i). We characterize the order parameters $L$ and $M$ of the AFM in $Mn_2Au$|Py by all-optical probes, both in the quasi-static and the ultrafast regime. We find that magnetic signal contributions dominate the optical response and that linear and quadratic probes give direct insight into antiferromagnetic spin dynamics, thereby resolving challenge (ii). In particular, we show that, with these probes, the NSOT torkance and the volume fractions of the four in-plane Néel-vector domains can be inferred. Therefore, once challenge (iii) is solved, the way to NSOT-driven ultrafast antiferromagnetic-order switching is open.

**II Experimental setup**

The schematic of our experimental approach is shown in Fig. 1a. In brief, our sample is a $Mn_2Au$|Py stack where the Py magnetization and, thus, Neel vector of the exchange-coupled $Mn_2Au$ is set by a quasi-static external magnetic field $B_{\text{ext}}$ that is varied quasi-statically as a function of time $t$.

To apply ultrafast NSOTs, a THz pump pulse is normally incident onto the sample stack. After a delay $\tau$, the sample state is interrogated by a normally incident and linearly polarized optical probe pulse (wavelength 800 nm, duration ~15 fs). We use the same probe pulses to also characterize the unperturbed sample in the absence of a pump pulse, in particular as a function of $B_{\text{ext}}$. In the following, we describe our experiment in more detail.

**II.A Sample and external magnetic field**

The samples consist of $Mn_2Au(001)(50\ nm)$ thin films grown epitaxially on $Ta(001)(13\ nm)|Mo(001)(20\ nm)$ double buffer layers on MgO(100) substrates by magnetron sputtering as described in detail in ref. [6]. A polycrystalline layer Py(10 or 6 nm) of Py (permalloy, $Ni_{80}Fe_{20}$) is deposited on top of the $Mn_2Au$ layer at room temperature, followed by a polycrystalline capping layer $SiN_x(2\ nm)$ to protect the metallic stack from oxidation. All measurements in the main text are performed on the sample with Py(10 nm). This sample has lateral dimensions of $10 \times 5\ mm^2$ and a nominal metal-film thickness of 93 nm. Its DC sheet resistance is determined as $1.2\ \Omega$ by a 4-point-probe measurement (Ossila Four-Point Probe), corresponding to an average conductivity of $8.8\ MS/m$.

To set the external magnetic field $B_{\text{ext}}$ parallel to the sample plane, we use two modes dedicated to the direction and amplitude of $B_{\text{ext}}$. In the first mode, we generate a static magnetic field $B_{\text{ext}}$ by a pair of permanent magnets with magnitude $B_{\text{ext0}} = 0.35\ T$. By mounting the magnets on a mechanical rotation stage, we can set the azimuthal angle $\varphi_{B_{\text{ext}}}$ of the magnetic field prior to each static or pump-probe measurement. Therefore, $\boldsymbol{B}_{\text{ext}}$ remains constant over the duration of the measurement and equals $\boldsymbol{B}_{\text{ext}} = B_{\text{ext0}}(\cos\varphi_{\boldsymbol{B}_{\text{ext}}} \boldsymbol{u}_x + \sin\varphi_{\boldsymbol{B}_{\text{ext}}} \boldsymbol{u}_y)$ (Fig. 2a).

In the second mode, an electromagnet generates a magnetic field $\boldsymbol{B}_{\text{ext}}(t) = B_{\text{ext}}(t)\boldsymbol{u}_x$ along $\boldsymbol{u}_x$. Its amplitude varies harmonically with time $t$ according to

$$B_{\text{ext}}(t) = -B_{\text{ext0}} \cos\frac{2\pi t}{2T}, \tag{1}$$

where the maximum field is $B_{\text{ext0}} = 0.19$ T. The period is chosen to be twice the time $T = 1$ ms between two consecutive pump pulses. This choice allows us to conduct pump-probe measurement for positive and negative magnetic field because $B_{\text{ext}}(t+T) = -B_{\text{ext}}(t)$.

**II.B Pump and probe details**

The THz pump and optical probe pulses are derived from an amplified Ti:sapphire laser system. It is fed by nanojoule-class pulses from a Ti:sapphire laser oscillator, resulting in millijoule-class pulses with an accordingly lower repetition rate. The pump pulses are derived from the amplified laser pulses (repetition rate $1/T = 1$ kHz), whereas the probe pulses are taken from part of the seed-laser beam (repetition rate $1/T_{\text{pr}} = 80$ MHz).

**II.B.1 Pump pulses**

The THz pump pulses are generated by excitation of a large-area Si-based spintronic THz emitter (STE) [7] with amplified laser pulses (pulse energy 7 mJ, center wavelength 800 nm, duration 40 fs). The pump-beam diameter of 4 cm (full width at half maximum of the intensity) results in a THz beam with a diameter of 2.8 cm. A 90° off-axis parabolic mirror focuses the THz beam onto the sample surface into a spot diameter of about 100 µm under normal incidence.

The transient focal incident THz electric field is measured by Zeeman-torque sampling (Figure S 1) [8] and amplitude calibration through a GaP window. We reach peak electric and magnetic fields of 1 MV/cm and 0.3 T, respectively.

The STE provides THz pulses with a short duration < 250 fs, which eases the separation of different signal contributions based on their potentially different time scales. As a further feature, we can accurately control the polarization plane and field polarity of the THz field by simply rotating the static magnetic field that sets the magnetization of the STE [7].

**II.B.2 Signal probing**

The probe beam is normally incident on the sample. After a probe has pulse has traversed the sample, we measure changes $S$ in the probe polarization, i.e., the rotation of the probe polarization plane and the ellipticity. To detect the probe rotation, we employ a combination of a half-wave plate, polarizing beam splitter and pair of balanced photodiodes. To detect the probe ellipticity, we convert ellipticity to rotation by introducing a quarter-wave plate before the half-wave plate.

We note that the signal $S$ is always subject to an unknown offset due, e.g., to static birefringence of the substrate or slow setup drifts. Therefore, only changes in $S$ due, e.g., to the pump pulse or variations of the external magnetic field, can be measured. The signal can be written as

$$S(t) = S_0(t) + \Delta S(t), \tag{2}$$

where $S_0(t)$ is the slowly varying signal due to quasi-static variations of $\boldsymbol{B}_{\text{ext}}$ vs probe time $t$. The pump-induced signal $\Delta S(t)$ typically varies on much shorter time scales due to the impact of the pump pulse.

Importantly, the probe pulses are derived from the seed oscillator of the laser system, which offers advantages relative to probe pulses from the amplified pulse train. First, they exhibit significantly reduced pulse-to-pulse variations and smaller pulse duration. Second, their about 80000 times higher repetition rate allows us to densely sample the signal $S(t)$ at times

$$t = t_{\text{pu}} + nT_{\text{pr}} + \tau. \tag{3}$$

Here, $t_{\text{pu}}$ is the arrival time of the pump. To detect the ultrafast pump-induced variations of signal $S(t)$, the fine delay $\tau$ can be set between $-400$ ps and $+400$ ps in steps of about 20 fs by a retroreflector and a mechanical translation stage that is traversed by the probe beam. Slower variations of $S(t)$ are sampled by the subsequent probe pulses that arrive at after delays of $nT_{\text{pr}}$, where $T_{\text{pr}} \approx T_{\text{pr}}/80000 = 12.5$ ns is the time between consecutive probe pulses, and $n$ is an integer.

To detect the photodiode signal $S(t)$ of each probe pulse [Eq. (3)] separately, we employ a sufficiently fast digitizer (National Instruments PCI-5122, sampling rate 100 MHz).

### II.B.3 Pump and probe timing

Fig. 1d and 1e schematically summarizes the timing of the experiment. As the signal $S(t)$ repeats periodically with period $2T$, we only need to consider the time interval from 0 to $2T$. First, the pump arrives periodically with period $T = 1$ ms, i.e., at times $t_{\text{pu}} = t_0$ and $t_{\text{pu}} = t_0 + T$, where $t_0$ is an offset.

Second, the amplitude $B_{\text{ext}}(t)$ of the external magnetic field $\boldsymbol{B}_{\text{ext}}(t) = B_{\text{ext}}(t)\boldsymbol{u}_x$ is cycled with a period of $2T$. Due to the doubled period time, consecutive pump pulses perturb the samples at exactly opposite magnetic fields because $B_{\text{ext}}(t + T) = -B_{\text{ext}}(t)$. Owing to the fast modulation of the magnetic field, we can measure the sample response to fields $\pm \boldsymbol{B}_{\text{ext}}$ with excellent signal-to-noise ratio. The offset $t_0$ is straightforwardly be controlled by an according phase shift of the periodic $B_{\text{ext}}(t)$.

Third, the probe samples $S(t)$ densely at times given by Eq. (3). Note that signal $S_0(t)$ is slow, i.e., $S_0(t + \tau) = S_0(t)$ to very good approximation because $B_{\text{ext}}(t + \tau) = B_{\text{ext}}(t)$ and $\tau \ll T$.

### II.C Signal phenomenology

The signals obtained for a normally incident probe pulse [4] obey the relationship

$$S \propto \langle P \rangle + \langle Q \rangle \cos(2\varphi_{\text{pr}}) + \langle R \rangle \sin(2\varphi_{\text{pr}}). \tag{4}$$

Each term with $P$, $Q$ and $R$ has a distinctly different dependence on the polarization-plane angle of the incident probe pulse (Fig. 1b). The angular bracket $\langle . \rangle$ denotes the average over the probed volume of the Mn$_2$Au|Py stack, which also takes the different probing sensitivity in Mn$_2$Au and Py into account, as summarized by Table 1. Therefore, for example, $\langle P \rangle$ is proportional to the volume-averaged local signal $P$. In the experiment, we focus on variations of $\langle P \rangle, \langle Q \rangle, \langle R \rangle$ induced by changes in the external magnetic field and/or the THz pump pulse.

In Eq. (4), the terms $P = P_S + P_N$, $Q = Q_S + Q_N$ and $R = R_S + R_N$ have contributions whose variations solely report on the changes in the spin degrees of freedom $S$, but also contributions due to variations of the non-spin degrees of freedom $N$ [4]. An expansion up to second order in $\boldsymbol{L}$ and $\boldsymbol{M}$ allows one to derive relationships between $P_S$, $Q_S$ and $R_S$ and the magnetic order parameters of the Mn$_2$Au|Py stack. Table 1 shows how the local $\boldsymbol{L}(\boldsymbol{r})$ and $\boldsymbol{M}(\boldsymbol{r})$ at position $\boldsymbol{r}$ contribute to the signal under the assumption that only the bulk properties of tetragonal Mn$_2$Au and isotropic Py need to be considered.

Changes in the components $P_N, Q_N, R_N$ can arise from, e.g., pump-induced modification of the electron and phonon occupations in the sample. Importantly, the pump-induced non-spin-related signal variations $\Delta P_N, \Delta Q_N, \Delta R_N$ cannot be separated from their spin counterparts $\Delta P_S, \Delta Q_S$ and $\Delta R_S$ based on symmetry arguments. For example, $\Delta P_N = \Delta P - \Delta P_S$ has the same dependence on the pump- and probe-field polarization and sample azimuth as $\Delta P_S$. Therefore, more sample-specific arguments need to be used to exclude sizeable non-spin contributions. For bare Mn$_2$Au, it was shown that the non-spin contributions are not dominant in $\langle Q \rangle$ [4].

Eq. (4) suggests the following strategy to extract $\langle P \rangle, \langle Q \rangle$ and $\langle R \rangle$: We measure the signal $S$ for at least 3 (typically 4 or 8) probe angles $\varphi_{\text{pr}}$, e.g., 0°, 45°, 90°. By Fourier transforming these data with respect to $\varphi_{\text{pr}}$, we can straightforwardly obtain $\langle P \rangle, \langle Q \rangle$ and $\langle R \rangle$ (Appendix A).

### III Results

First, we magneto-optically characterize sample vs quasi-static variation of $\boldsymbol{B}_{\text{ext}}$. Next, we conduct THz-pump optical-probe measurements, permitting calibration of the magneto-optic signals with respect to $\boldsymbol{M}$ and $\boldsymbol{L}$. Finally, we use the obtained information to determine the volume fractions of the 0°, 90°, 180° and 270° domains.

In the following, $\boldsymbol{L}$ and $\boldsymbol{M}$ refer to the Neel vector and magnetization in the AFM, if not mentioned otherwise, whereas $\boldsymbol{M}^{\text{FM}}$ refers to the FM.

## III.A Quasi-static variation of $B_{\text{ext}}$

To characterize the static magnetic properties of the sample, we quasi-statically vary the direction and amplitude of the applied static in-plane magnetic field $\boldsymbol{B}_{\text{ext}}$ and simultaneously measure the probe signal $S_0$.

We assume that variation of $\boldsymbol{B}_{\text{ext}}$ does not change the non-spin signals $P_\mathcal{N}$, $Q_\mathcal{N}$, $R_\mathcal{N}$. Therefore, any signal variation reports on changes in $P_S$, $Q_S$, $R_S$ and, thus, in the magnetic order parameters of the sample stack according to Table 1.

### III.A.1 Variation of $B_{\text{ext}}$ direction

We start with measuring the probe rotation $S_0$ vs the azimuthal angle $\varphi_{\boldsymbol{B}_{\text{ext}}}$ of $\boldsymbol{B}_{\text{ext}}$ with $B_{\text{ext0}} = 0.35$ T for 4 linear probe polarizations (Section II.A and Figure S 2). By means of Eq. (4) (see Appendix A), we obtain the signal contributions $\langle P_0 \rangle$, $\langle Q_0 \rangle$ and $\langle R_0 \rangle$, which are displayed in Fig.2a.

The major signal variation vs $\varphi_{\boldsymbol{B}_{\text{ext}}}$ arises from component $\langle R_0 \rangle$ of Eq. (4) (Fig.2a). A smaller contribution by $\langle Q_0 \rangle$ is due potentially to a rotational offset of approximately 6° between the probe polarization plane and the sample easy axis, which projects a small part of the dominant $\langle R_0 \rangle$ part onto $\langle Q_0 \rangle$. Finally, component $\langle P_0 \rangle$ exhibits a minor dependence on $\varphi_{\boldsymbol{B}_{\text{ext}}}$ (Fig.2a).

To understand the measured $\langle R_0 \rangle$ vs $\varphi_{\boldsymbol{B}_{\text{ext}}}$, we first note that $\langle R_0 \rangle$ equals the weighted spatial integral over $L_{\|0}^2 \cos(2\varphi_L)$ and $M_{\|0}^2 \cos(2\varphi_M)$ (Table 1). As Mn$_2$Au is a strong easy-plane antiferromagnet [9], we assume that $L_{\|0}^2 \approx L_0^2$. Second, according to previous magnetometry studies [6], the coercive field of the Mn$_2$Au|Py sample studied here is expected to be below 0.1 T. Therefore, saturation can be assumed in our measurement, provided $\boldsymbol{B}_{\text{ext}}$ is along an easy axis of Mn$_2$Au [6]. Consequently, $\langle R_0 \rangle$ vs $\varphi_{\boldsymbol{B}_{\text{ext}}}$ should vary symmetrically around zero. This property allows us to fix the unknown offset of the measured $\langle R_0 \rangle$ vs $\varphi_{\boldsymbol{B}_{\text{ext}}}$, as already done in Fig.2a.

Finally, the signal $\langle R_0 \rangle$ has the same magnitude for Py(6 nm) and Py(10 nm) capping layers (Figure S 2) and, therefore, originates from Mn$_2$Au rather than Py. This notion is consistent with the fact that the Mn$_2$Au layer is 5 times thicker than the Py layer for the sample shown in Fig.2. Likewise, reflection measurements on prealigned Mn$_2$Au films [10] suggest that MLB in this material is significantly larger than in Py [11].

To summarize, the measured $\langle R_0 \rangle$ in Fig.2a predominantly stems from Mn$_2$Au and scales like

$$\langle R_0 \rangle = r_L L_0^2 \langle \cos(2\varphi_L) \rangle, \tag{5}$$

where the sensitivity $r_L$ is assumed to be homogeneous over the probed AFM volume. The dependence of $\langle R_0 \rangle$ on $\varphi_{\boldsymbol{B}_{\text{ext}}}$ indicates the following points. First, the Néel-vector orientation can be set along any of the 4 easy axes by the external field, as indicated by the schematic in Fig.2a. Second, the signal changes abruptly between the saturated states at $\varphi_{\boldsymbol{B}_{\text{ext}}} \approx 0°, 90°, 180°, 270°$, virtually within the experimental step size of 15°. Therefore, most domains in the sampled volume rotate at once into the next easy axis.

Note that, according to Eq. (5), saturation of $\langle R_0 \rangle$ for $\boldsymbol{B}_{\text{ext}}$ along $\boldsymbol{u}_x$ or $\boldsymbol{u}_y$ implies, respectively, $\langle \cos(2\varphi_L) \rangle|_{\text{sat}} = \pm 1$ and, thus, signals of $\pm r_L L_0^2$. The signals for opposite directions of $\boldsymbol{L}$ are always equal. From Fig.2a and $\varphi_{\boldsymbol{B}_{\text{ext}}} = \varphi_L = 0$, we infer

$$\langle R_0 \rangle|_{\text{sat}} = r_L L_0^2 \approx 2.2 \text{ mrad} \tag{6}$$

for the probe rotation and also ellipticity in magnetic saturation (sat) (Figure S 2). We, thus, obtain the first magneto-optic coefficient for the studied thin film, which is consistent with previous reflection-configuration studies on prealigned Mn$_2$Au films [10]. Its value can also be found in Table 3 and is comparable to the linear longitudinal magneto-optic-Kerr-effect signal in Fe [12] and exceeds the quadratic magneto-optic Kerr response of ferromagnetic Py [11] and Fe [13] thin films by about one order of magnitude.

### III.A.2 Variation of $B_{\text{ext}}$ amplitude.

Next, we harmonically vary the external field amplitude $B_{\text{ext}}$ along the $x$ direction $\boldsymbol{u}_x$, i.e., $\boldsymbol{B}_{\text{ext}} = B_{\text{ext}}(t)\boldsymbol{u}_x$, from $B_{\text{ext}}(0) = -B_{\text{ext0}}$ to $+B_{\text{ext}}(T) = 0.19$ T and back (Fig.2b and Section II.B.2). To obtain a high signal-to-noise ratio, one cycle is swept within the period $2T = 2$ ms and repeated over 10 runs.

Fig 2b shows the signal contributions $\langle P_0 \rangle$, $\langle Q_0 \rangle$ and $\langle R_0 \rangle$ vs time $t$ along with $B_{\text{ext}}(t)$. First, as with the data in Fig.2a and Eq. (5), $\langle R_0 \rangle$ reports on the weighted spatial integral $L_0^2 \langle \cos(2\varphi_L) \rangle$, which equals $L_0^2$ in saturation. Consequently, to fix the unknown signal offset in $\langle R_0 \rangle$ vs $B_{\text{ext}}$, we assume that, at the maximum field of $B_{\text{ext0}}$, the signal saturates and, thus, equals the peak signals in Fig.2a, where $B_{\text{ext0}} = 0.35$ T. This assumption is supported by the shape of the signal measured here and the coercive field of $< 0.1$ T found by slower measurements in Ref. [6].

Second, the signal minimum of $\langle R_0 \rangle \approx 0$ at $B_{\text{ext}} \approx 0.1$ T implies $\langle \cos(2\varphi_L) \rangle \approx 0$ and, equivalently, $\langle L_x^2 \rangle \approx \langle L_y^2 \rangle$ [Eq. (5) and Table 1]. In other words, the volume fractions of $x$ and $y$ domains are approximately equal. Finally, the signal minimum of $\langle R_0 \rangle \approx 0$ implies a coercive field of $B_c \approx 0.1$ T. This value is slightly larger than in previous findings on similar samples [6] and may be due the faster variation of the external magnetic applied here [14].

Our conclusions about the antiferromagnetic order are summarized by the schematic at the bottom of Fig.2b. Note that this schematic assumes that the volume of domains with $\boldsymbol{L}_0 \upuparrows \boldsymbol{u}_x$ and $\boldsymbol{L}_0 \upuparrows -\boldsymbol{u}_x$ is equal. We will determine the precise domain volume fractions in Section III.C.

### III.B Pump-probe signals

Following a magneto-optic characterization of our Mn$_2$Au|Py sample as a function of a quasi-static external magnetic field $\boldsymbol{B}_{\text{ext}}$ (Fig.2), we now excite it with a normally incident THz electromagnetic field $F = (\boldsymbol{E}, \boldsymbol{B})$ (Fig.1). The pump pulse induces a variation of the probe signal by $\Delta S(t) = S(t) - S_0(t)$, which depends on the delay $\tau = t - t_{\text{pu}}$ between pump and probe pulse.

To separate signal contributions to $\Delta S(t)$ linear and quadratic in $F$, we use pump fields $\pm F$ of opposite polarity. Here, reversal of the THz field is straightforwardly implemented with our spintronic THz high-field source [7]. To better separate spin- and non-spin-related signals, we also reverse the polarity of the external magnetic field $\boldsymbol{B}_{\text{ext}} = B_{\text{ext}}\boldsymbol{u}_x$ analogous to Fig.2b, which switches $B_{\text{ext}}$ between $\pm 0.19$ T at every other pump pulse.

### III.B.1 Expected signal contributions and temporal responses

According to Eq. (4), $\Delta S(t)$ arises from pump-induced changes $\langle \Delta P \rangle$, $\langle \Delta Q \rangle$, $\langle \Delta R \rangle$. In contrast to the quasi-static case, the signals may not only have contributions $\Delta P_S$, $\Delta Q_S$, $\Delta R_S$ related exclusively to spin dynamics but also contributions $\Delta P_N$, $\Delta Q_N$, $\Delta R_N$ reporting on changes in non-spin degree of freedom.

**True pump-induced spin dynamics**

We start with mechanisms of true pump-induced spin dynamics and the signal they result in.

***Linear responses.*** To facilitate the further discussion, we first focus on the most common types of spin dynamics in linear response to a terahertz electromagnetic field. Symbolically, the linear response of the component $Y$ of a magnetic order parameter to a component of the driving THz field $F = (\boldsymbol{E}, \boldsymbol{B})$ can be written as

$$\Delta Y(t) = (H_{\Delta YX} * X)(t) = \int dt' \, H_{\Delta YX}(t - t') X(t'). \tag{7}$$

Here, $H(t)$ is the response $\Delta Y(t)$ to an impulsive driving field component $X(t') = \delta(t')$ at time $t' = 0$. **Table 2** and Fig. 3 summarize possible $X$, $\Delta Y$ and $H_{\Delta YX}$ mathematically and graphically, respectively.

First, the field-like Zeeman torque $\propto \boldsymbol{M} \times \boldsymbol{B}$ by the magnetic field $\boldsymbol{B}$ of the THz pulse perturbs the magnetization $\boldsymbol{M}^{\text{FM}}$ of the ferromagnetic layer FM=Py [12, 15, 16]. This resulting transient out-of-plane magnetization component $\Delta M_z^{\text{Py}}$ may be detected through $\langle \Delta P \rangle$ [Eq. (4)]. The subsequent precession is so slow that $H_{\Delta YX}$ is a step function on the picosecond time scale (Table 2 and Fig. 3).

The Zeeman torque may also excite dynamics in the antiferromagnetic Mn$_2$Au layer [17]. In our sample geometry, we expect an out-of-plane contribution $\Delta L_z$ as well as an in-plane magnetization $\Delta M_y$ [4]. From Raman scattering experiments [2] and theory [3], a frequency of the order of about 3-4 THz is expected. However, the signal contribution from this mode can be neglected in our geometry as we lack sensitivity to $\Delta L_z$ and the net $\Delta M_y$ is expected to be very small.

Second, in Mn$_2$Au, field-like Néel spin-orbit torques (NSOTs) by the THz electric field $\boldsymbol{E}$ leads to in-plane rotation of $\boldsymbol{L}$ by an angle $\Delta\varphi_L$ and to an out-of-plane $\Delta M_z$ [4]. The latter can straightforwardly be explained by the macrospin $\sigma$-model for antiferromagnets [18-20], which implies that the dynamic magnetization $\Delta\boldsymbol{M}^{\mathrm{AFM}}$ of an antiferromagnet with $|\Delta\boldsymbol{M}^{\mathrm{AFM}}| \ll |\boldsymbol{L}|$ follows the Neel vector $\boldsymbol{L}$ through

$$\Delta\boldsymbol{M}^{\mathrm{AFM}} \approx \frac{1}{\gamma B_{\mathrm{ex}} L_0} \boldsymbol{L} \times \partial_t \boldsymbol{L}. \tag{8}$$

In Mn$_2$Au, the in-plane rotation of the Neel vector by $\Delta\varphi_L$ is resonant and related to a magnon (Fig.1c) [4] whose oscillatory motion is found in Table 2 and Fig. 3. If the $\boldsymbol{L}$ motion is dominated by $\Delta\varphi_L$, Eq. (8) results in $\Delta M_z^{\mathrm{AFM}} \propto L_0^2 \partial_t \Delta\varphi_L$, and the response functions are related by a time derivative as well (Table 2 and Fig. 3).

***Quadratic responses.*** While Zeeman torque and NSOTs can already induce changes in $\boldsymbol{M}$ and $\boldsymbol{L}$ to linear order in the driving field $F$, there can also be changes that scale with the square of $F$ to lowest order. An important example is ultrafast quenching of magnetic order by terahertz pulses [21]. In our probing geometry and for the layer AFM, a reduction of $|\boldsymbol{L}|$ could be probed by the spin-related signal contribution $\Delta R_S^+ \propto \Delta(L_x^2) \approx 2 L_{0x} \Delta L_x$ (Table 1).

**Non-spin contributions**

The signal contributions $\Delta S_\mathcal{N}$ that, at least partially, report on the dynamics of non-spin degrees of freedom $\mathcal{N}$ can be classified in (i) symmetry-conserving and (ii) symmetry-breaking variations of $\mathcal{N}$.

A possible microscopic scenario of (i) is an isotropic increase of the energy of electrons or phonons in AFM or FM. It scales with the energy density $\propto \boldsymbol{E}^2$ of the driving THz field to lowest order. Because the sample symmetry is not lowered in this process, the resulting signal changes have the same structure as the $P_S$, $Q_S$ and $R_S$ in Table 1. However, as the spin degrees of freedom remain unchanged, $\mathcal{S} = \mathcal{S}_0$, the signal is mediated by variations of the magneto-optic coefficients, such as $r_L$ [21-23]. Therefore, the signal $\Delta R$ has a non-spin-related component of the form $\Delta R_\mathcal{N} = L_{\parallel 0}^2 \cos(2\varphi_{L_0}) \Delta r_L$, which cannot be distinguished from pure spin dynamics using symmetry arguments.

Even more challenging, effects quadratic in $\boldsymbol{E}$ are of even order in $\boldsymbol{L}_0$ and can, therefore, not be separated from contributions of nonmagnetic sample parts by reversal of $\boldsymbol{L}_0$. An example is the THz Kerr effect [24, 25] in the isotropic substrate of our sample.

A possible microscopic mechanism leading to (ii) is a $\boldsymbol{E}$-induced shift of the Fermi sphere of AFM or FM. The shift scales with $\boldsymbol{E}$ to lowest order, is accompanied by a net electron current and transiently lowers the sample symmetry. Phenomenologically, it is most suitably described in explicit terms of the driving field [4]. For example, for $\Delta Q$, the respective contribution can be written as $\Delta Q_\mathcal{N} \propto (L_{0x} E_x - L_{0y} E_y)$. Therefore, even though this specific $\Delta Q_\mathcal{N}$ reverses as $\boldsymbol{L}_0$ is reversed, it cannot be distinguished from its truly spin-related counterpart based on symmetry properties.

Therefore, in the following, we will make use of more specific arguments for the spin-dominant nature of the measured signals.

**III.B.2 Raw-signal analysis**

Typical pump-probe signals $\Delta S(t)$ are shown in Figure S 3 and Figure S 4 for various probe-polarization angles $\varphi_{\mathrm{pr}}$ (Fig.1b). Both rotation and ellipticity signals are acquired, as noted in the text where applicable. We extract the signal contributions $\langle\Delta P\rangle$, $\langle\Delta Q\rangle$, $\langle\Delta R\rangle$ of Eq. (4) based on their probe polarization dependence, as detailed in Section II.C and Appendix A. All contributions are shown in Figure S 5.

At first glance, signals that stem exclusively from spin dynamics are expected to exhibit the same temporal dynamics in both rotation (rot) and ellipticity (ell), e.g., $\langle\Delta P(\tau)\rangle|_{\mathrm{rot}} \propto \langle\Delta P(\tau)\rangle|_{\mathrm{ell}}$. Note, however, that different contributions, such as Zeeman torque and NSOTs, may contribute with different relative weight to $\langle \Delta P(\tau) \rangle|_{\rm rot}$ and $\langle \Delta P(\tau) \rangle|_{\rm ell}$, thus, cause different dynamics of these two observables. Consequently, in the following, we will first focus on signals that can be well explained by a single signal contribution from Table 1.

We start with signals odd in both the driving field $F$ and the quasi-static external magnetic $\boldsymbol{B}_{\rm ext}$ (Fig.3) and subsequently consider signals even in both $F$ and $\boldsymbol{B}_{\rm ext}$ (Fig.4). In other words, odd and even signals $\Delta S^{\pm}$ are given by

$$\Delta S^{\pm} = \frac{\Delta S(+F, +\boldsymbol{B}_{\rm ext}) \pm \Delta S(-F, +\boldsymbol{B}_{\rm ext}) \pm \Delta S(+F, -\boldsymbol{B}_{\rm ext}) + \Delta S(-F, -\boldsymbol{B}_{\rm ext})}{4}, \quad (9)$$

and, in an analogous manner, their odd and even components $\langle \Delta P^{\pm} \rangle$, $\langle \Delta Q^{\pm} \rangle$, $\langle \Delta R^{\pm} \rangle$.

### III.B.3 Odd signals and interpretation

As seen in Figure S 5, the dominant components of the probe ellipticity signal $\Delta S^{-}$ odd in $F$ and $\boldsymbol{B}_{\rm ext}$ are $\langle \Delta P^{-} \rangle$ and $\langle \Delta Q^{-} \rangle$. They are shown in Fig.4a along with the measured driving electric field. Fig.3b confirms that the probe ellipticity signal $\Delta S^{-}$ odd in $F$ and $\boldsymbol{B}_{\rm ext}$ scales linearly with the THz field $F$. In other words, the signals in Fig.4a characterize the linear-response regime. Once the driving THz field has decayed, both $\langle \Delta P^{-} \rangle$ and $\langle \Delta Q^{-} \rangle$ feature an oscillatory pattern (Fig.3a), which is reminiscent of the expected response functions shown in Table 2 and Fig.3.

However, for the probe rotation signals, we observe more complex dynamics with additional contributions, which can be attributed to further processes, as we discuss below (Figure S 5, Figure S 13). Therefore, we here focus on the ellipticity signals.

According to Tables 1 and 2, the MCB component $\langle \Delta P \rangle$ contains a term due to the dynamic out-of-plane magnetization $M_z$, while the MLB component $\langle \Delta Q \rangle$ contains terms due to the in-plane Néel vector through $2\Delta(L_x L_y) \approx 2L_{\|0}^2 \Delta\varphi_L$ and the in-plane magnetization through $2\Delta(M_x M_y)$. The occurrence of dynamic variations of $\varphi_L$ and $M_z$ linear in $F, \boldsymbol{B}_{\rm ext}$ is consistent with NSOT-driven with the homogeneous in-plane magnon of Mn$_2$Au [4]. Accordingly, we fit the signal as $\Delta\varphi_L \propto H_{\Delta\varphi_L E_x} * E_x$ with the impulse response function $H_{\Delta\varphi_L E_x}$ (Table 2) and obtain a frequency of $\Omega/2\pi = 0.6$ THz and damping rate $\Gamma = 2.3$ ps$^{-1}$ in good agreement with previous results [4]. The excellent agreement between data and fit supports the assignment

$$\langle \Delta Q^{-} \rangle = 2q_L L_0^2 \langle \Delta\varphi_L \rangle, \quad (10)$$

where we used $L_{\|0} = L_0$. Fig.3c further reveals that the calculated time derivative $\partial_t \Delta\langle Q^{-} \rangle$ largely agrees with $\Delta\langle P^{-} \rangle$. This behavior is expected from the $\sigma$-model [Eq. (8)] and indicates that

$$\langle \Delta P^{-} \rangle = p_{\boldsymbol{M}} \langle \Delta M_z \rangle. \quad (11)$$

As $\Delta \boldsymbol{M}^{\rm FM}$ is expected to follow a different response function (Table 2), we conclude that it is not efficiently probed in the ellipticity signal.

We note that the signal assignments by Eqs. (10) and (11) for Mn$_2$Au|Py are fully consistent with our previous findings on prealigned Mn$_2$Au films. In these samples, the smaller contribution $\propto \Delta M_z$ could not be detected in $\langle \Delta P^{-} \rangle$, presumably due to the smaller signal-to-noise ratio.

The odd signal $\langle \Delta Q^{-} \rangle$ well characterizes the temporal dynamics of $\langle \Delta\varphi_L \rangle$. However, it cannot determine the absolute magnitude of the deflection $\langle \Delta\varphi_L \rangle$. A static calibration of the necessary magneto-optical coefficient $q_L L_0^2$ as done in Section III.A.1 for $r_L L_0^2$, is not possible since the static signal $\langle Q_0 \rangle$ vanishes. We, thus, look for an even pump-probe contribution $\langle \Delta R^{+} \rangle$ that could be calibrated by a nonvanishing $\langle R_0 \rangle$.

### III.B.4 Even signals

Consequently, we turn to the signal $\Delta S^{+}$ even in $F$ and $\boldsymbol{B}_{\rm ext}$. As discussed in Sections III.B.1-2, $\Delta S^{+}$ as well can have a variety of non-spin-related contributions $\Delta S_{\mathcal{N}}^{+}$. They are even more challenging to identify than in $\Delta S^{-}$ because, by definition, $\Delta S^{-}$ changes sign as, e.g., $\boldsymbol{L}_0$ is reversed, whereas $\Delta S^{+}$ remains unchanged. Therefore, $\Delta S^{+}$ can have contributions that are completely unrelated to $\boldsymbol{L}_0$ of the sample.

Fig.4a shows component $\langle \Delta R^+ \rangle$ for both rotation and ellipticity signals. Fig.4b confirms the signals of Fig.4a scale with the square $F^2$ of the driving field, which is the lowest order even in $F$.

Note that the ellipticity- and rotation-type $\langle \Delta R^+ \rangle$ coincide once the instantaneous pump intensity $\boldsymbol{E}^2(\tau)$ becomes negligible ($\tau > 0.2\,\text{ps}$, Fig.4a). In the ellipticity signal, an additional sharp feature is present close to the overlap of pump and probe pulses ($\tau \approx 0\,\text{ps}$), which we attribute to a non-magnetic refractive-index change of the metallic film. Importantly, the otherwise identical dynamics of the ellipticity- and rotation-type $\langle \Delta R^+ \rangle$ suggests that both signals report on the same component of spin dynamics. An even signal $\Delta R_\mathcal{N}$ from the MgO substrate is possible, but would be box-shaped and extend over a much longer delay than 0.2 ps [25] and can, thus, be discarded here.

According to Table 1, the spin-related contribution to $\langle \Delta R^+ \rangle$ is a superposition of $\Delta \langle L_{0\parallel}^2 \cos(2\varphi_L) \rangle$ and $\Delta \langle M_{0\parallel}^2 \cos(2\varphi_M) \rangle$, i.e., a change in (i) the in-plane magnitudes ($L_{0\parallel}$, $M_{0\parallel}$) and/or (ii) rotation ($\varphi_L$, $\varphi_M$) of the order parameters $\boldsymbol{L}$, $\boldsymbol{M}$. Assuming scenario (ii) prevails, the signal would scale according to $\langle \Delta R^+ \rangle \propto \Delta \langle \cos(2\varphi_L) \rangle \approx 2\langle (\Delta \varphi_L)^2 \rangle$ and, thus,

$$\langle \Delta R^+ \rangle = 2 r_L L_0^2 \langle (\Delta \varphi_L)^2 \rangle. \tag{12}$$

In this case, Eqs. (10) and (12) imply that the signal $\langle \Delta R^+ \rangle$ would scale with the square of signal $\langle \Delta Q^- \rangle$. Indeed, the direct comparison of $\langle \Delta Q^- \rangle$ (Fig.3a) and $\langle \Delta R^+ \rangle$ (Fig.4a) in Fig.4c confirms that, approximately, $\langle \Delta R^+(t) \rangle \propto \langle \Delta Q^-(t) \rangle^2$. This remarkable agreement indicates that scenario (i), i.e., ultrafast magnetic-order quenching by pump-induced heating of Mn$_2$Au and Py, makes a minor contribution to $\langle \Delta R^+ \rangle$.

We note that the signal $\langle \Delta R^+(t) \rangle \propto [\Delta \varphi_L(t)]^2$ displays ultrafast dynamics with twice the frequency of $\langle \Delta Q^-(t) \rangle \propto \Delta \varphi_L(t)$, which is consistent with similar observations in the insulating antiferromagnet NiO [26].

We emphasize that the observed pump-induced changes in the Neel vector and signal $\langle \Delta Q^- \rangle$ scale linearly with the pump field $F$ (Fig.3b), whereas the quadratic scaling of $\langle \Delta R^+ \rangle$ (Fig.3b) is just a consequence of the probe process [Table 1, Eqs. (10) and (12)]. Thus, our findings indicate that we probe the in-plane rotation $\Delta \varphi_L$ of the Neel vector of Mn$_2$Au through both $\langle \Delta Q^- \rangle$ and $\langle \Delta R^+ \rangle$ in a complementary manner.

### III.B.5 Absolute deflection angle and magneto-optic coefficients

As summarized by Eq. (12), the even signal $\langle \Delta R^+ \rangle$ is found to probe the in-plane $\boldsymbol{L}$ deflection (Section III.B.4). This observable has a measurable nonvanishing quasi-static analog $\langle R_0 \rangle$, which can be used to calibrate the pump-induced signal $\langle \Delta R^+ \rangle$ and, thus, the magnitude of the transient deflection angle $\Delta \varphi_L$.

More precisely, we assume that the sample is in the magnetically saturated state and acknowledge that the probe spot is much smaller than the THz pump spot. Consequently, the Neel vector $\boldsymbol{L}_0$ and the pump electric field $\boldsymbol{E}$ are homogeneous over the probed sample volume, resulting in $\langle \cos(2\varphi_L) \rangle = 1$ and, thus, $\langle R_0 \rangle = r_L L_0^2$ for the quasi-static case, and $\langle (\Delta \varphi_L)^2 \rangle = (\Delta \varphi_L)^2$ and, thus, $\langle \Delta R^+ \rangle = -2 r_L L_0^2 (\Delta \varphi_L)^2$ for the pump-probe signal. Therefore, we obtain the absolute $(\Delta \varphi_L)^2$ by taking

$$\left| \frac{\langle \Delta R^+ \rangle}{\langle R_0 \rangle} \right|\bigg|_{\text{sat}} = 2(\Delta \varphi_L)^2, \tag{13}$$

which determines the proportionality coefficient $\langle (\Delta \varphi_L)^2 \rangle / \langle \Delta R^+ \rangle$ of Eq. (12) in magnetic saturation (sat). Importantly, knowledge of $(\Delta \varphi_L)^2$ allows us to also calibrate the proportionality coefficient $\langle \Delta \varphi_L \rangle / \langle \Delta Q^- \rangle$ of Eq. (10) by using

$$\sqrt{\langle \Delta Q^- \rangle^2 \left| \frac{\langle R_0 \rangle}{\langle \Delta R^+ \rangle} \right|}\bigg|_{\text{sat}} = |q_L L_0^2| \tag{14}$$

Note that Eq. (10) is valid for all pump-probe delays, whereas Eqs. (12)-(14) are accurate for pump-probe delays $\tau > 0.3\,\text{ps}$, i.e., after the pump pulse has decayed (Section III.B.4). We obtain a peak deflection of $\Delta \varphi_L = -5°$ at $\tau = 0.13\,\text{ps}$. As shown in Figure S 13, $q_L L_0^2$ is about 6 times smaller in rotation than in ellipticity.

With $\Delta\varphi_L$ and Eq. (8), we further determine the relative transient magnetization in AFM by $\Delta M_z/L_0 = \Delta\varphi_L/\gamma B_{\text{ex}}$. With $B_{\text{ex}} = 1300$ T [9] and the gyromagnetic ratio $\gamma = 0.176\ \text{T}^{-1}\text{ps}^{-1}$ of the electron, we find that $\Delta M_z/L_0$ reaches a maximum value of $2.7 \cdot 10^{-3}$. It is interesting to use this value to estimate the maximum MCB signal change (Table 1) due to a hypothetical spin-flip transition in AFM, where both spin sublattices finally align parallel to $\boldsymbol{u}_z$, i.e., $\Delta M_z/L_0 = 1$. We obtain 16 mrad (ellipticity) and 8 mrad (rotation, Figure S 13). Thus, the MCB signal due to $\Delta M_z = L_0$ is about one order of magnitude larger than the maximum attainable MLB signal. However, in our pump-probe experiment, the induced MCB signals are smaller than the MLB signals due to the small AFM net magnetization.

Finally, by estimating the THz electric field strength inside layer AFM with the Tinkham formula [27] and an average conductivity of $8.8\ \text{MS/m}$, we determine a torkance of $100\ \text{cm}^2\text{A}^{-1}\text{s}^{-1}$ (Appendix D). We emphasize this value is in good agreement with our previous estimate based on the onset of non-linear magnon dynamics [4].

### III.B.6 Intermediate summary

From the quasi-static variation of the azimuthal angle for the external magnetic field $\boldsymbol{B}_{\text{ext}}$ (Fig. 2a), we can obtain the magneto-optic coefficient $r_L L_0^2$ [Eq. (5)]. The variation of the amplitude of $\boldsymbol{B}_{\text{ext}}$ (Fig. 2b) induces quasi-static switching of AFM that can be probed by signal contribution $\langle R_0 \rangle$.

Our THz-pump optical-probe experiments show responses linear (Fig. 4) and quadratic (Fig. 5) in the driving THz field. We find that the dynamics of AFM are probed by three distinct signal contributions reporting on $\langle \Delta P^- \rangle \propto \Delta M_z$ [Eq. (11)], $\langle \Delta Q^- \rangle \propto \langle \Delta\varphi_L \rangle$ [Eq. (10)] and $\langle \Delta R^- \rangle \propto \langle (\Delta\varphi_L)^2 \rangle$ [Eq. (12)]. The dynamics can be explained consistently by the antiferromagnetic in-plane magnon mode, yet the different signal contributions provide complementary insights into the dynamics.

Together with the static amplitude $\langle R_0 \rangle$, these results allow us to quantify the absolute local transient Neel-vector deflection $\Delta\varphi_L$, reaching up to 5°, which corresponds to a torkance of $100\ \text{cm}^2\text{A}^{-1}\text{s}^{-1}$. Further, we can extract all magneto-optic coefficients of layer AFM as $p_M L_0 = (8 + 16\text{i})$ mrad, $q_L L_0^2 = (0.3 + 1.8\text{i})$ mrad and $r_L L_0^2 = (2.2 + 2.2\text{i})$ mrad, where the real and imaginary part denotes, respectively, probe rotation and ellipticity (Table 3). Importantly, with our method, we can calibrate the magneto-optic coefficient $q_L L_0^2$, which is only accessible in dynamic, but not static processes.

### III.C Revealing the domain distribution

So far, we have focused on the saturated, i.e., virtually uniform, magnetic state of the sample. In a non-saturated state, we expect different scaling of the contributions $\langle \Delta Q^- \rangle$ and $\langle \Delta R^+ \rangle$, which can give unique insights into the domain distribution, which is also highly interesting for ultrafast switching experiments. As a proof of concept, we probe the non-saturated sample state during the quasi-static switching in an external field, as shown in Fig. 6a, which is analogous to Fig. 2b.

In general, there are 2 in-plane easy axes for the Néel vector and, consequently, 4 distinct directions. Thus, a multi-domain state can be described by 4 fractions

$$f_{\varphi_{L_0}} = \frac{V_{\varphi_{L_0}}}{V}, \tag{15}$$

of the total probed volume. Here, $V_{\varphi_{L_0}}$ is the volume with angle $\varphi_{L_0} = 0°, 90°, 180°, 270°$ of $\boldsymbol{L}$ relative to the easy direction $\boldsymbol{u}_x$. The fractions fulfill

$$\sum_{\varphi_{L_0}} f_{\varphi_{L_0}} = 1. \tag{16}$$

### III.C.1 Scaling of quasi-static signals

For a multi-domain state within the probed volume, the quasi-static signal is through Eq. (5) given by

$$\langle R_0 \rangle = r_L L_0^2 \langle \cos(2\varphi_{L_0}) \rangle = r_L L_0^2 \sum_{\varphi_{L_0}} f_{\varphi_{L_0}} \cos(2\varphi_{L_0}), \tag{17}$$

Because we have $\cos(2\varphi_{L_0}) = +1$ for horizontal domains ($\varphi_{L_0} = 0°, 180°$), $\cos(2\varphi_{L_0}) = -1$ for vertical domains ($\varphi_{L_0} = 90°, 270°$) and, by Eq. (16), $f_{0°} + f_{90°} + f_{180°} + f_{270°} = 1$, Eq. (17) becomes

$$\langle R_0 \rangle = r_L L_0^2 [2(f_{0°} + f_{180°}) - 1] \tag{18}$$

As indicated in the schematic of Fig. 6a, for saturation at $B_{\text{ext}}(t) = -B_{\text{ext}0}$, we can assume $f_{180°} = 1$ and, likewise, $f_{0°} = 1$, at $B_{\text{ext}}(t) = +B_{\text{ext}0}$. Both cases yield a signal $\langle R_0 \rangle = r_L L_0^2$, whereas at times at times $t$ during switching, one may expect a roughly equal distribution among the 4 orientations.

To facilitate an elegant comparison with the pump-probe signals in the following, we further separate the signal $\langle R_0 \rangle$ into components

$$\langle R_0^{\pm}(\boldsymbol{B}_{\text{ext}}) \rangle := \frac{\langle R_0(\boldsymbol{B}_{\text{ext}}) \rangle \pm \langle R_0(-\boldsymbol{B}_{\text{ext}}) \rangle}{2} \tag{19}$$

that are even ($R_0^+$) and odd ($R_0^-$) in the external field $\boldsymbol{B}_{\text{ext}}$. Note that $\langle R_0^+ \rangle \approx \langle R_0 \rangle$, while $\langle R_0^- \rangle \approx 0$, i.e., the quasi-static signal is predominantly even in the external magnetic field. Therefore, we can rearrange Eq. (18) as

$$\frac{1}{2}\left(1 \pm \frac{\langle R_0^+ \rangle}{r_L L_0^2 V}\right) = \begin{cases} f_{0°} + f_{180°} \\ f_{90°} + f_{270°} \end{cases}. \tag{20}$$

The domain volume fractions parallel to the $x$ axis ($f_{0°} + f_{180°}$) and the $y$ axis ($f_{90°} + f_{270°}$) as extracted by Eq. (20) are shown in Fig.6b vs $B_{\text{ext}}/B_{\text{ext}0}$. Because the MCB signal $R_0$ is quadratic in $\boldsymbol{L}_0$, it cannot distinguish between domains with Néel vector $\pm \boldsymbol{L}_0$, as also implied by Eq. (20). Therefore, no statement about the exact domain configuration is possible. To mitigate this lack of information, we add THz-pump optical-probe signals at various times $t$ to our analysis.

### III.C.2 Pump-probe measurements for various signal states

To probe the sample at various states during the cycle of the quasi-statically varying magnetic field $\boldsymbol{B}_{\text{ext}}(t) = B_{\text{ext}}(t)\boldsymbol{u}_x$, we perform a pump-probe experiment at various times during this cycle, which has a duration of $2T$. As summarized in Section II.B.3 and Fig. 1d and 1e, the THz pump pulse arrives at times $t_{\text{pu}} = t_0$ and $t_{\text{pu}} = t_0 + T$, where $B_{\text{ext}}(t_0)$ and $B_{\text{ext}}(t_0 + T) = -B_{\text{ext}}(t_0)$ have opposite values. Note that the external magnetic field is static for the pump-probe delay $\tau \ll T$ used in our experiment, i.e., $B_{\text{ext}}(t_{\text{pu}} + \tau) = B_{\text{ext}}(t_{\text{pu}})$.

To reduce measurement time and probe the magnetic state vs applied field $B_{\text{ext}}(t)$ with fine $t$ resolution, we measure only the minimum number of signals necessary to separate all contributions. We do not reverse the electric field, either, since signals odd in only either $F$ or $\boldsymbol{B}_{\text{ext}}$ are negligible on the time scales of the antiferromagnetic mode (Figure S 3, Figure S 4). As detailed in Appendix C, for $\varphi_{\text{pr}} = 45°$ and rotation signals, we use the approximation

$$\langle \Delta R^+ \rangle \approx \frac{\Delta S(F, +B_{\text{ext}}) + \Delta S(F, -B_{\text{ext}})}{2}, \tag{21}$$

while, for ellipticity signals that change sign between $\varphi_{\text{pr}} = 0°$ and $\varphi_{\text{pr}} = 90°$, we employ

$$\langle \Delta Q^- \rangle \approx \frac{\Delta S(F, +B_{\text{ext}}) - \Delta S(F, -B_{\text{ext}})}{2}. \tag{22}$$

The opposite field of $B_{\text{ext}}(t)$ is simply obtained by considering the signal at time $t + T$ where $B_{\text{ext}}(t + T) = -B_{\text{ext}}(t)$.

### III.C.3 Scaling of pump-probe signals

Field-like NSOTs are proportional to the product $\boldsymbol{L} \cdot \boldsymbol{E}$ [4, 28] and, thus, $\boldsymbol{L}_0 \cdot \boldsymbol{E}$ in the linear-response limit considered here. Therefore, within a single $\boldsymbol{L}_0$ domain, the resulting deflection of the Néel vector scales according to $\Delta \varphi_L \propto \boldsymbol{L}_0 \cdot \boldsymbol{E} \propto \cos \varphi_{L_0} E_x + \sin \varphi_{L_0} E_y$. Therefore, upon reversing $\boldsymbol{L}_0 \to -\boldsymbol{L}_0$, $\Delta \varphi_L \to -\Delta \varphi_L$, reverses, too [4], and $\Delta \varphi_L$ vanishes if $\boldsymbol{L}_0 \perp \boldsymbol{E}$ [4, 28].

**Signal $\langle \Delta R^+ \rangle$.** As a consequence, the pump-probe signal $\langle \Delta R^+ \rangle$ [Eq. (12)] is expected to follow

$$\langle \Delta R^+ \rangle = 2r_L L_0^2 \left[(f_{0°} + f_{180°})(\Delta \varphi_L^x)^2 - (f_{90°} + f_{270°})(\Delta \varphi_L^y)^2\right], \tag{23}$$

where $\Delta\varphi_L^j(t) = (H_{\Delta\varphi_{LE_j}} * E_j)(t)$ is the time-dependent deflection of the Néel vector by the THz electric-field component $E_j$ with $j = x, y$ (Table 2). Eq. (23) implies that the signal amplitude for the two orthogonal pump fields from Eq. (23) should scale according to

$$\langle \Delta R^+ \rangle \propto \begin{cases} +(f_{0°} + f_{180°}) & \text{for } \mathbf{E} \parallel \mathbf{u}_x \\ -(f_{90°} + f_{270°}) & \text{for } \mathbf{E} \parallel \mathbf{u}_y. \end{cases} \quad (24)$$

The proportionality factor in Eq. (24) can be simply determined taking the signal $\langle \Delta R^+ \rangle|_{\text{sat}}$ for $\mathbf{E} \parallel \mathbf{u}_x$ in the magnetically saturated state (sat), i.e., $B_{\text{ext}} = B_{\text{ext0}}$, where $f_{0°} \approx 1$ or $f_{180°} \approx 1$.

The amplitudes of the measured traces $\langle \Delta R^+(\tau) \rangle$ are determined by a projection method (Appendix B and Figure S 7). The resulting relative amplitude $\langle \Delta R^+ \rangle / \langle \Delta R^+ \rangle|_{\text{sat}}$ with respect to the signal at maximum external field $B_{\text{ext0}}$ vs $B_{\text{ext}}/B_{\text{ext0}}$ are shown in Fig. 6b. We find that $\langle \Delta R^+ \rangle$ for $\mathbf{E} \parallel \mathbf{u}_x$ and $f_{0°} + f_{180°}$ as obtained through static measurements [Section III.C.1 and Eq. (20)] agree excellently. Likewise, Fig. 6b confirms that, for $\mathbf{E} \parallel \mathbf{u}_y$, $\langle \Delta R^+ \rangle \propto f_{90°} + f_{270°}$, in agreement with Eq. (24). We conclude that $\langle \Delta R^+ \rangle$ is dominated by spin degrees of freedom, and $\langle R_0^+ \rangle$ and $\langle \Delta R^+ \rangle$ report consistently on the domain statistics. However, neither of the signals is able to distinguish $\pm \mathbf{L}_0$ domains.

**Signal $\langle \Delta Q^- \rangle$.** Consequently, we consider the signal contribution $\langle \Delta Q^- \rangle$. By explicitly performing the volume averaging in Eq. (10), we obtain the relationship

$$\langle \Delta Q^- \rangle = 2 q_L L_0^2 \left[ (f_{0°} - f_{180°}) \Delta\varphi_L^x - (f_{90°} - f_{270°}) \Delta\varphi_L^y \right] \quad (25)$$

and, thus

$$\langle \Delta Q^- \rangle \propto \begin{cases} +(f_{0°} - f_{180°}) & \text{for } \mathbf{E} \parallel \mathbf{u}_x \\ -(f_{90°} - f_{270°}) & \text{for } \mathbf{E} \parallel \mathbf{u}_y. \end{cases} \quad (26)$$

Therefore, both $\langle \Delta Q^- \rangle$ [Eq. (26)] and $\langle \Delta R^+ \rangle$ [Eq. (24)] report on the temporal dynamics of $\Delta\varphi_L$, but the dependence on the quasi-static domain distribution is complementary and scales like $f_{0°} \pm f_{180°}$ and $f_{90°} \pm f_{270°}$. Note that the minus sign in front of $f_{180°}$ in Eq. (26) arises because $\Delta\varphi_L$ reverses as $\mathbf{L}_0$ reverses.

Fig. 6c displays the signal amplitude of $\langle \Delta Q^- \rangle$ (see Figure S 8). For $\mathbf{E} \parallel \mathbf{u}_x$, the zero-crossing of $\langle \Delta Q^- \rangle$ indicates switching when $B_{\text{ext}} \approx 0.5 B_{\text{ext0}}$, consistent with the behavior observed in $\langle R_0^+ \rangle$ and $\langle \Delta R^+ \rangle$ (Fig. 6b). The quasi-static MCB from Py (Figure S 6) indicates that $\mathbf{M}^{\text{FM}}$ and $\mathbf{L}$ reverse simultaneously. For $B_{\text{ext}} > 0.5 B_{\text{ext0}}$, a signal $\langle \Delta Q^- \rangle \propto f_{90°} - f_{270°} \neq 0$ is observed (Fig. 6c). For $\mathbf{E} \parallel \mathbf{u}_y$, $\langle \Delta Q^- \rangle$ is non-zero even in the saturated state ($B_{\text{ext}} \approx B_{\text{ext0}}$, Fig. 6b), which indicates that $\mathbf{E}$ is not exactly perpendicular to the antiferromagnetic easy axis and, thus, exerts NSOTs.

### III.C.4 Full domain statistics

To estimate the volume fractions $f_{0°}, f_{90°}, f_{180°}, f_{270°}$ of the 4 domains at any time $t$ and, thus, $B_{\text{ext}}(t)$, we use Eq. (16). and, thus, we only require 3 independent observables from our measurements of $\langle R_0^+ \rangle$ and $\langle \Delta Q^- \rangle$ with $\mathbf{E} \parallel \mathbf{u}_x$ and $\mathbf{E} \parallel \mathbf{u}_y$, respectively (see Appendix E). The individual domain fractions are subsequently obtained by linear combination of the observables.

Fig. 6d shows the extracted domain volume fractions vs external field $B_{\text{ext}}/B_{\text{ext0}}$, where the data are interpolated on a finer grid for better visibility. For $B_{\text{ext}} = -0.8 B_{\text{ext0}}$, virtually all domains are at 180° with respect to $\mathbf{u}_x$, i.e., $f_{180°} = 1$. Our extraction procedure of $f_{0°}, f_{90°}, f_{180°}, f_{270°}$ is subject to systematic uncertainties due to imperfect alignment of $\mathbf{E}$, which, in our case manifests itself in a small negative fraction $f_{90°}$ at $B_{\text{ext}} = -0.8 B_{\text{ext0}}$. Upon increasing $B_{\text{ext}}$ toward zero, the fraction of 0°, 90° and 270° starts growing. Interestingly, at $B_{\text{ext}} = 0.5 B_{\text{ext0}}$, a small dominance of the 90° and 270° domains can be observed, i.e., $f_{90°} + f_{270°} > f_{0°} + f_{180°}$. Thus, the expected distribution sketched in Fig. 2b is largely reproduced, but slightly modified.

### IV Summary and discussion

**Signals are dominated by spins.** We measure magneto-optic signals for quasi-static and ultrafast variations of the magnetic order of an AFM|FM exchange-spring system. Importantly, all signals predominantly report on true spin dynamics, including signals even in $\mathbf{L}_0$. More precisely, the observed magnetic linear birefringence (MLB) signals report on the dynamic Néel vector deflection $\Delta\varphi_L$ in linear and quadratic order, i.e., $\langle \Delta Q^-(t) \rangle \propto \Delta\varphi_L(t)$ and $\langle \Delta R^+(t) \rangle \propto [\Delta\varphi_L(t)]^2$. Therefore, the observed signals $\langle \Delta R^+(t) \rangle$ quadratic in the driving THz field result exclusively from the probing mechanism while the spin dynamics $\Delta\varphi_L(t)$ are still in the linear-response regime. The complementary information content of $\langle \Delta Q^-(t) \rangle$ and $\langle \Delta R^+(t) \rangle$ allow for the identification of pure spin dynamics and even calibration of the amplitude of $\Delta\varphi_L(t)$. In this context, we note that MLB signals, as opposed to MCB, are appropriate tools to probe the antiferromagnetic dynamics in Mn$_2$Au, as they yield larger signals in typical situations where $|\Delta \boldsymbol{M}| \ll |\boldsymbol{L}_0|$.

To summarize, the ultrafast pump-induced signals have precisely the same interpretation [Eq. (5)] as the signals for quasi-static variation of $\boldsymbol{L}$, $\boldsymbol{M}$ by an external magnetic field [Eqs. (5), (10), (11), (12)]. This analogy indicates that the pump-induced signals can be transferred from the small-signal response to the general response as with the quasi-static case, i.e.,

$$\langle \Delta Q^- \rangle \propto L_0^2 \langle \Delta \sin(2\varphi_L) \rangle \quad \text{and} \quad \langle \Delta R^+ \rangle \propto L_0^2 \langle \Delta \cos(2\varphi_L) \rangle. \tag{27}$$

The complementary information of the dynamics probed by $\langle \Delta Q^- \rangle$ and $\langle \Delta R^+ \rangle$ is expected to become essential in the highly non-linear and, ultimately, ultrafast switching regime, where $\langle \Delta \sin(2\varphi_L) \rangle$ does not distinguish between 0°, 90°, 180° and 270° domains, whereas $\langle \Delta \cos(2\varphi_L) \rangle$ does.

**Spin torques and temporal dynamics.** The observed spin dynamics are very well explained by the in-plane antiferromagnetic magnon mode within the $\sigma$-model, which dictates a strict relationship between the dynamics of the Néel vector and the net magnetization [Eq. (8)]. The mode is driven predominantly by Néel spin-orbit torques and the amplitude calibration allows for the extraction of the torkance in good agreement with previous results [4].

While these results indicate that the exchange-spring system preserves the general characteristics of the antiferromagnetic spin dynamics, we note that an even better fit of the data in Fig. 4a can be obtained by a superposition of two sinusoidal modes as a response (Figure S 9). This fact may indicate the presence of a (non-homogeneous) standing spin wave in AFM due to the exchange-spring behavior. A related effect was observed in the spin dynamics of the Py layer by ferromagnetic resonance experiments [1].

**Other signal contributions.** In our presentation of results, we have focused largely on results that can be explained entirely by the antiferromagnetic dynamics. Here, we address the remaining features of the dynamic response.

First, there is a fast feature in the ellipticity-type $\langle \Delta R^+(t) \rangle$ that resembles the square $\boldsymbol{E}^2(t)$ of the THz field. We attribute this component to an anisotropic electron distribution in momentum space of AFM or FM, which relaxes with the momentum relaxation time of ~10 fs [27].

Second, the MCB signal $\langle \Delta P^- \rangle$ is expected to also contain a contribution due the change $\Delta M_z^{FM}$ in the Py magnetization, driven by off-resonant Zeeman torque [12]. The rotation part of $\langle \Delta P^- \rangle$ contains a contribution that is consistent with this scenario (Figure S 13).

Third, in the rotation signal $\langle \Delta Q^- \rangle$ (Figure S 13), we find an additional fast component that follows the temporal waveform of the THz pump field.

**Domain statistics during switching.** An interesting feature of the quasi-static reversal shown in Fig.6b is that both quasi-static and pump-probe signals consistently indicate an imbalance $f_{90°} + f_{270°} > f_{0°} + f_{180°}$ at $B_{\text{ext}} = 0.5 B_{\text{ext0}}$. As the extraction methods are independent of each other, we conclude that this imbalance is not due to a systematic offset in either method.

There is a second imbalance $f_{90°} > f_{270°}$ at $B_{\text{ext}} = 0.5 B_{\text{ext0}}$ (Fig.6d). The origin of this peculiar feature cannot be directly determined from our results, but it may indicate that local strains modify the anisotropy to favor 90° vs 270° orientations [29, 30]. Domain-wall motion in the reversal may play a further important role in this context.

## V Conclusion

To summarize, we show that AFM|FM exchange-spring systems serve as good model systems for antiferromagnetic spintronics, providing external control of the Néel vector $\boldsymbol{L}$ without significant alteration of the ultrafast dynamics. We stress that a single macroscopic $\boldsymbol{L}$-domain can only be formed in

exchange-spring systems where the antiferromagnet terminates with a single sublattice at the interface [6]. A recent theoretical study indicates that this property occurs in a broad range of materials [31]. Therefore, our methods is very useful for studying other antiferromagnetic exchange-spring systems.

Our results also show that THz pulses allow one to implement of NSOTs-based probing that is sensitive to the direction of $L$ and not only the orientation of $L^2$, in contrast to commonly used techniques, like XMLD [32]. More generally, our experiment is a special realization of a second-order optical response of the sample. It is, thus, similar to other non-linear optical techniques, such as second-harmonic generation[33], but explicitly takes advantage of resonant THz magnon excitation.

For the potential study of ultrafast coherent switching of $L$ in $Mn_2Au$, we have shown that challenges (i), i.e., external control of $L$, and (ii), i.e., exploration and calibration of magneto-optic responses of $Mn_2Au$, are now met. We are, thus, left with the final step of (iii) increasing the NSOTs magnitude, e.g. by employing high-field terahertz sources [34] or resonant microstructure antennas [35, 36].

## Appendices

### Appendix A: Signal phenomenology and separation

The probe-signal phenomenology as given by Eq. (4) is obtained as described in ref. [4] [Eq. (9)] for normal incidence of the probe pulse with a polarization angle $\varphi_{\mathrm{pr}} = \sphericalangle(\boldsymbol{E}_{\mathrm{pr}}, \boldsymbol{u}_x)$. Here, $\boldsymbol{E}_{\mathrm{pr}}$ is the probe electric field, and $\boldsymbol{u}_x$ is a crystallographic high symmetry direction of Mn$_2$Au. To obtain Table 1, we assume the spatial symmetries of Mn$_2$Au are described by the tetragonal point group 4/mmm and that Py can be treated as an isotropic material in the absence of magnetization. We further neglect the symmetry breaking at surfaces and interfaces.

To extract the MCB contribution $\langle P \rangle$ and the MLB components $\langle Q \rangle$ and $\langle R \rangle$, we note that the functions $\cos(2\varphi_{\mathrm{pr}})$, $\sin(2\varphi_{\mathrm{pr}})$ and $1 = \cos(0\varphi_{\mathrm{pr}})$ in Eq. (4) are mutually orthogonal. Consequently, we measure the signals $S(t, n)$ for a set of equidistant probe-polarization angles $\varphi_{\mathrm{pr},n} = 2\pi n/N$ in the interval $[0, \pi[$, for $n = 0, \ldots, N/2 - 1$, where $N$ is even and $N/2 \geq 4$. We then assume that for $n' = N/2, \ldots, N-1$, we can write $S(t, n') = S(t, n' - N)$, i.e., the signals are identical for $\varphi_{\mathrm{pr}} \to \varphi_{\mathrm{pr}} + \pi$.

Because $2\cos(2\varphi_{\mathrm{pr}}) = \mathrm{e}^{\mathrm{i}2\varphi_{\mathrm{pr}}} + \mathrm{e}^{-\mathrm{i}2\varphi_{\mathrm{pr}}}$, $2\mathrm{i}\sin(2\varphi_{\mathrm{pr}}) = \mathrm{e}^{\mathrm{i}2\varphi_{\mathrm{pr}}} - \mathrm{e}^{-\mathrm{i}2\varphi_{\mathrm{pr}}}$ and $1 = \mathrm{e}^{\mathrm{i}0\varphi_{\mathrm{pr}}}$, the $\langle Q(t) \rangle$, $\langle R(t) \rangle$ and $\langle P(t) \rangle$ are, respectively, given by $[\tilde{s}(t, 2) + \tilde{s}(t, -2)] = 2\,\mathrm{Re}\,\tilde{s}(t, 2)$, $[\tilde{s}(t, 2) - \tilde{s}(t, -2)]/\mathrm{i} = 2\,\mathrm{Im}\,\tilde{s}(t, 2)$ and $\tilde{s}(t, 0)$. Here,

$$\tilde{s}(t, k) = \frac{1}{N} \sum_{n=0}^{N-1} S(t, n)\,\mathrm{e}^{-2\pi \mathrm{i} n k / N} \tag{28}$$

is the discrete Fourier transformation of $S(t, n)$ with respect to $n$ at integer frequency $k = -\frac{N}{2}, \ldots, \frac{N}{2} - 1$, which has an angle period of $2\pi/k$. For the quasi-static signals in Fig.2a, $N/2 = 4$ polarizations are used (Figure S 2), while for the signals in Fig.2b (Figure S 12) and Fig.4a and Fig.5a, $N/2 = 8$ polarizations are used (Figure S 3-S5).

### Appendix B: Extraction of signal amplitude vs. reference signal

To obtain the relative amplitude $A$ of a signal waveform $\Delta S(\tau_k)$, measured at discrete time delays $\tau_k$ with $k = 0, \ldots, N - 1$, with respect to a reference signal $\Delta S_{\mathrm{ref}}(\tau_k)$ on the same time grid, we simply project $\Delta S$ on $\Delta S_{\mathrm{ref}}$ by the linear correlation factor [37]

$$A = \frac{\sum_{k=0}^{N-1} \Delta S(\tau_k) \Delta S_{\mathrm{ref}}(\tau_k)}{\sum_{k=0}^{N-1} \Delta S_{\mathrm{ref}}(\tau_k) \Delta S_{\mathrm{ref}}(\tau_k)}. \tag{29}$$

Importantly, this method is linear with respect to $\Delta S(\tau_k)$ and, thus, conserves the relative sign of $\Delta S(\tau_k)$. Second, it accounts for all sampled points, as opposed to the, e.g. the maximum value. Finally, it is less sensitive to signal contributions that do not follow the temporal dynamics of the reference $\Delta S_{\mathrm{ref}}(\tau_k)$.

### Appendix C: Signals odd and even in external magnetic field

The signal contributions with respect to the external magnetic field and can be separated as $\Delta S = \Delta S^+_{\boldsymbol{B}_{\mathrm{ext}}} + \Delta S^-_{\boldsymbol{B}_{\mathrm{ext}}}$. Here,

$$\Delta S^{\pm}_{\boldsymbol{B}_{\mathrm{ext}}} := \frac{\Delta S(F, +\boldsymbol{B}_{\mathrm{ext}}) \pm \Delta S(F, -\boldsymbol{B}_{\mathrm{ext}})}{2} \tag{30}$$

is the component even (+) and odd (−) in the magnetic field.

For probe polarization $\varphi_{\mathrm{pr}} = 45°$, the signal is given by $\Delta S = \langle \Delta P \rangle + \langle \Delta R \rangle$ [Eq. (4)]. As shown in Figure S 3, signals odd in only either $F$ or $\boldsymbol{B}_{\mathrm{ext}}$ are negligible, $\langle \Delta P \rangle$ is predominantly odd in $\boldsymbol{B}_{\mathrm{ext}}$, and $\langle \Delta R \rangle$ is predominantly even. Thus, for $\varphi_{\mathrm{pr}} = 45°$, we have $\Delta S^-_{\boldsymbol{B}_{\mathrm{ext}}} \approx \langle \Delta P^- \rangle$ and $\Delta S^+_{\boldsymbol{B}_{\mathrm{ext}}} \approx \langle \Delta R^+ \rangle$.

When probing with $\varphi_{\mathrm{pr}} = 0°$ and $90°$, the signals are $\langle \Delta P \rangle + \langle \Delta Q \rangle$ and $\langle \Delta P \rangle - \langle \Delta Q \rangle$, respectively. Thus, the signal contribution changing sign is $\langle \Delta Q \rangle$. As shown in Figure S 3, $\langle \Delta Q \rangle$ does not contain contributions odd in $\boldsymbol{B}_{\mathrm{ext}}$ and even in $F$. Therefore, we have $\langle \Delta Q^-_{\boldsymbol{B}_{\mathrm{ext}}} \rangle \approx \langle \Delta Q^- \rangle$.

## Appendix D: Calibration of torkance

In the frequency domain, the amplitude $E$ of the linearly polarized THz electric field inside the film between two homogeneous half-spaces is given by the thin-film formula [27]

$$\frac{E}{E_{\text{inc}}} = \frac{2n_1}{n_1 + n_2 + Z_0 G}. \tag{31}$$

Here, $E_{\text{inc}}$ is the incident field, $n_1$ and $n_2$ is the refractive index of the first half-space (air, $n_1 = 1$) and the second half-space (MgO, $n_2 \approx 3$), respectively. $Z_0$ is the vacuum impedance and $G$ is the sheet conductance of the metal film.

In the time domain, the linear response of the Néel vector to $E$ is determined by $\Delta\varphi_L = A H_{\Delta\varphi_L E_x} * E$, where the impulse response function $H_{\Delta\varphi_L E_x}$ is given in Table 2. The coefficient $A$ depends on the torkance $\lambda_{\text{NSOT}}$ [4, 28] through

$$A = \gamma B_{\text{ex}} \sigma \lambda_{\text{NSOT}}. \tag{32}$$

We use the fit result shown in Fig.3a, the amplitude calibration (Section III.B.5), the measured $\sigma = 8.8\,\text{MS}\,\text{m}^{-1}$ and $G = \sigma d$, where $d = 93$ nm is the total metal-film thickness. We obtain $A \approx 2.1 \cdot 10^7\,\text{m}\,\text{V}^{-1}\text{s}^{-1}$, which yields $\lambda_{\text{NSOT}} \approx 100\,\text{cm}^2\,\text{A}^{-1}\text{s}^{-1}$, where the literature value of $B_{\text{ex}} = 1300$ T [9] is used.

## Appendix E: Extraction of domain statistics

To obtain the domain volume fractions shown in Fig. 6d, we combine various observables as follows. First, the domain contribution $\alpha := f_{0°} + f_{180°}$ is obtained from the quasi-static data $\langle R_0^+ \rangle$ according to Eq. (20). Second, $\beta := f_{0°} - f_{180°}$ and $\gamma := f_{90°} - f_{270°}$ are obtained from the pump-probe signals $\langle \Delta Q^- \rangle / \langle \Delta Q^- \rangle|_{\text{sat}}$ for $\boldsymbol{E} \parallel \boldsymbol{u}_x$ and $\boldsymbol{E} \parallel \boldsymbol{u}_y$, respectively [Eq. (26)].

It follows that $f_{0°} = (\alpha + \beta)/2$, $f_{180°} = (\alpha - \beta)/2$, $f_{90°} = (1 - \alpha + \gamma)/2$ and, finally, $f_{270°} = 1 - (f_{0°} + f_{90°} + f_{270°})$.

**Supplementary Information**

**Figures**

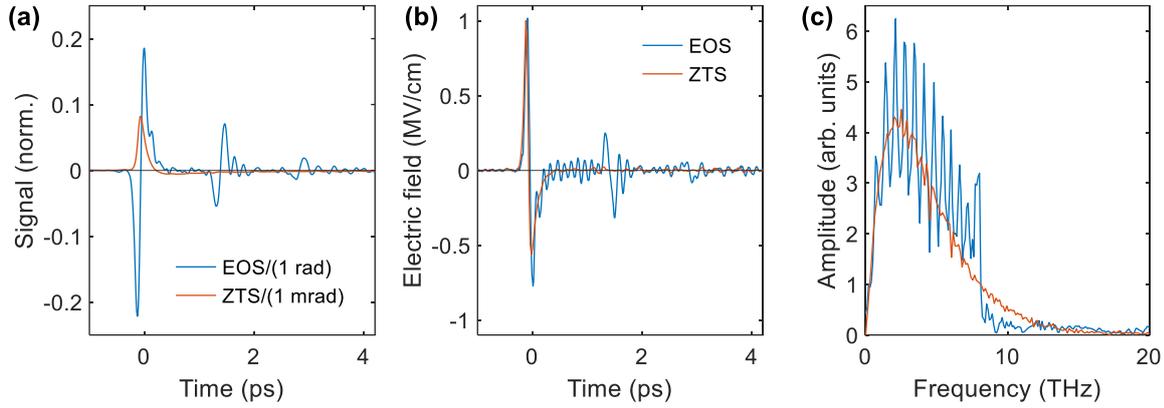

Figure S 1: THz electric field characterization. **(a)** Probe ellipticity signals from electro-optic sampling (EOS) in a GaP window with thickness 50 µm (blue line) and Zeeman-torque sampling (ZTS) from a Fe film with thickness 8 nm (orange line) [12]. The replicas in the EOS signal are due to reflections inside the detector. **(b)** Extracted THz electric field in air from the signals in panel (a). The replicas are not accounted for in the extraction procedure, but are absent in air. The amplitude of the ZTS waveform was calibrated by the low frequency amplitude of the EOS signal. **(c)** Fourier transformation of the electric-field waveforms in panel (b). The modulation of the EOS spectrum is due to the replicas.

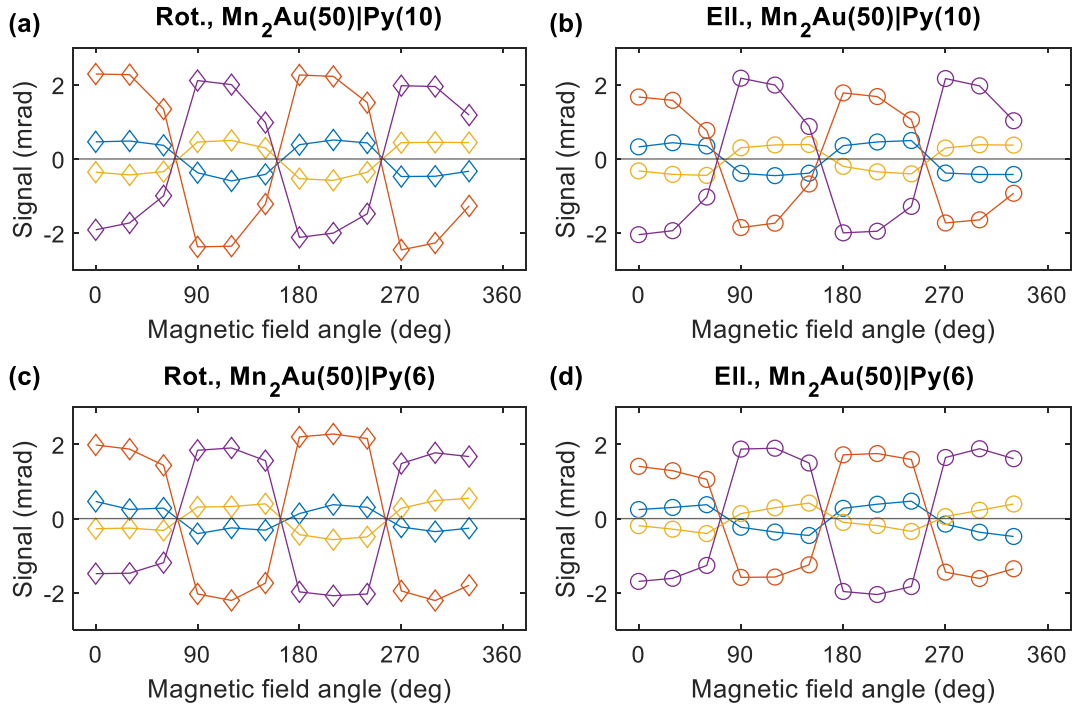

Figure S 2: Signals for rotating magnetic field $\varphi_{B_{\text{ext}}}$ in rotation (diamonds) and ellipticity (circles) for two Py-layer thicknesses of **(a,b)** 10 nm **(c,d)** 6 nm, where $\varphi_{\text{pr}} = 0°$ ($E_{\text{pr}} \parallel u_x$, blue symbols), $\varphi_{\text{pr}} = 45°$ (orange), $\varphi_{\text{pr}} = 90°$ (yellow), $\varphi_{\text{pr}} = 135°$ (purple).

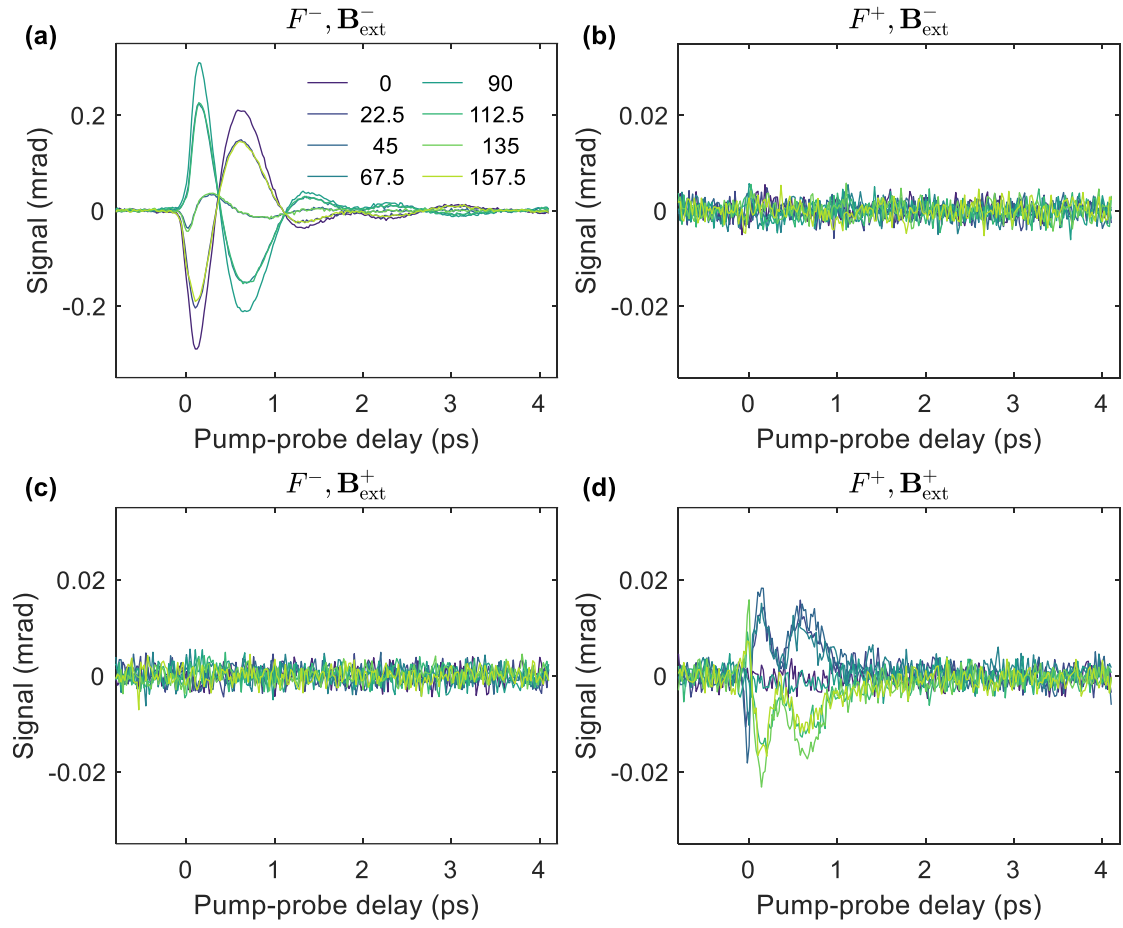

Figure S 3: Pump-probe ellipticity signals $\Delta S(\tau)$ for 8 probe polarizations, where $\varphi_{\mathrm{pr}} = \sphericalangle(\boldsymbol{E}_{\mathrm{pr}}, \boldsymbol{u}_x)$. Signals **(a)** odd in both THz field $F$ and external magnetic field $\boldsymbol{B}_{\mathrm{ext}}$, **(b)** even in $F$, odd in $\boldsymbol{B}_{\mathrm{ext}}$, **(c)** odd in $F$, even in $\boldsymbol{B}_{\mathrm{ext}}$ and **(d)** even in both $F$ and $\boldsymbol{B}_{\mathrm{ext}}$.

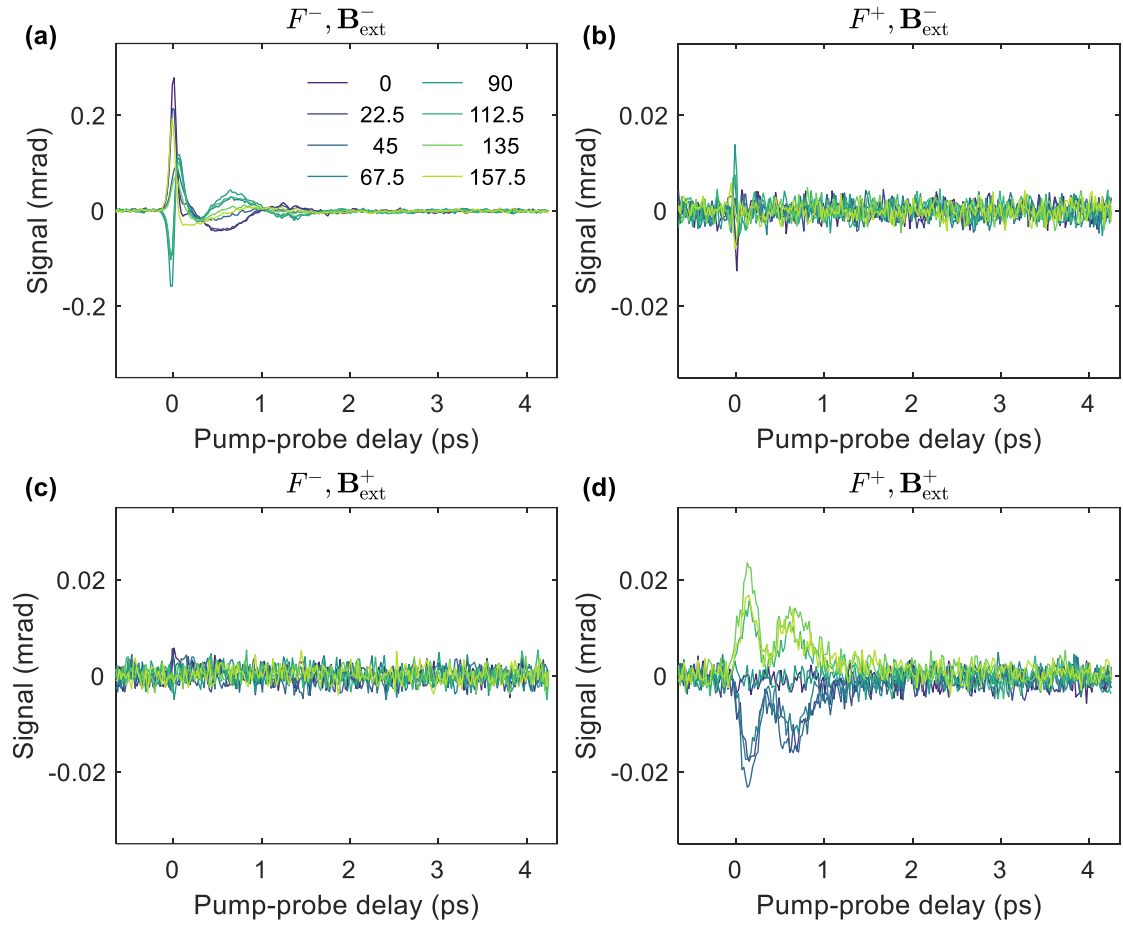

Figure S 4: Pump-probe rotation signals $\Delta S(\tau)$ for 8 probe polarizations, where $\varphi_{\mathrm{pr}} = \sphericalangle(\boldsymbol{E}_{\mathrm{pr}}, \boldsymbol{u}_x)$ Signals **(a)** odd in both THz field $F$ and external field $\boldsymbol{B}_{\mathrm{ext}}$, **(b)** even in $F$, odd in $\boldsymbol{B}_{\mathrm{ext}}$, **(c)** odd in $F$, even in $\boldsymbol{B}_{\mathrm{ext}}$ and **(d)** even in both $F$ and $\boldsymbol{B}_{\mathrm{ext}}$.

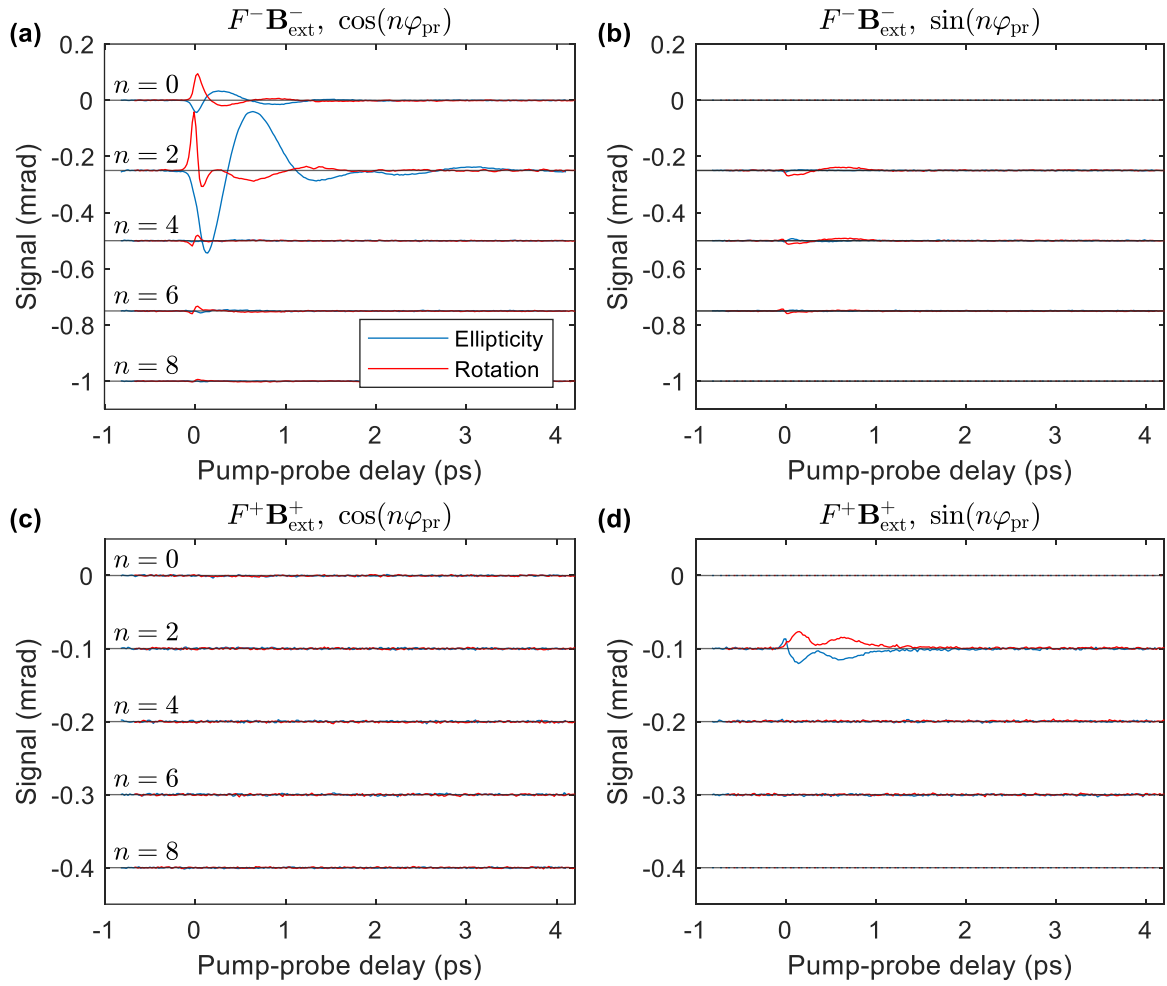

Figure S 5: $n$-fold signal contributions for the signals shown in Figure S 3 (ellipticity) and Figure S 4 (rotation). Blue lines are ellipticity signals, red lines the rotation signals. Traces are offset vertically for clarity. **(a, b)** Contributions $F^- \mathbf{B}_{\text{ext}}^-$ (odd in both) and **(c,d)** $F^+ \mathbf{B}_{\text{ext}}^+$ (even in both). Other contributions are much smaller (see Figure S 3 and S4) and not shown.

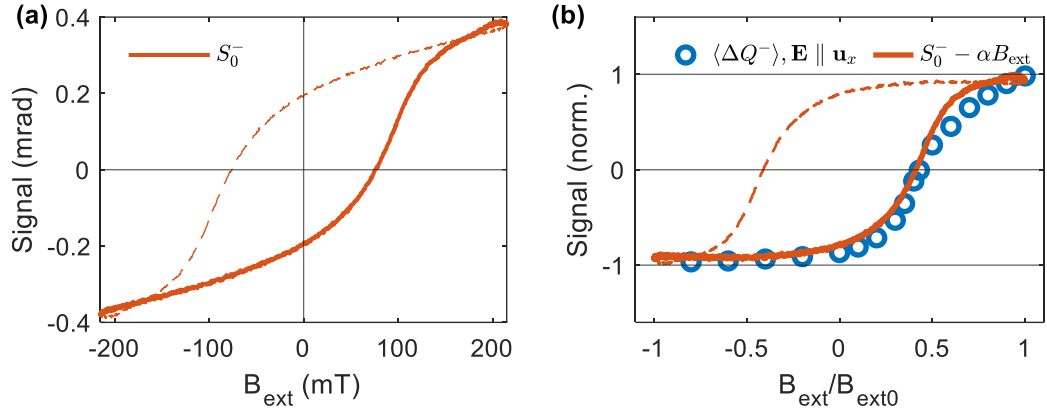

Figure S 6: **(a)** Ellipticity $S_0^-$ signal odd in external field $\boldsymbol{B}_{\text{ext}}$ vs $\boldsymbol{B}_{\text{ext}} = B_{\text{ext}}\boldsymbol{u}_x$. In contrast to the other experiments shown in this work, the angle of incidence is set to 30° to access the in-plane magnetization with the Faraday effect (MCB). The solid line is the signal for increasing $B_{\text{ext}}$, dashed line for decreasing $B_{\text{ext}}$. The signal is attributed to MCB in the Permalloy layer, i.e., $S_0^- \propto \boldsymbol{M}^{\text{FM}} \cdot \boldsymbol{u}_x$, since it vanishes for normal incidence. Signal contributions $\propto \boldsymbol{L}$ are not allowed by symmetry, and the quasi-static net magnetization of Mn$_2$Au is negligible. The linear trend is attributed to a Faraday-effect contribution $\propto B_{\text{ext}}$ in the MgO substrate. **(b)** Pump-probe signals (blue circles) from Fig. 5b for THz electric field along $\boldsymbol{u}_x$. The orange line is the signal from panel (a) where the linear trend was fitted and removed.

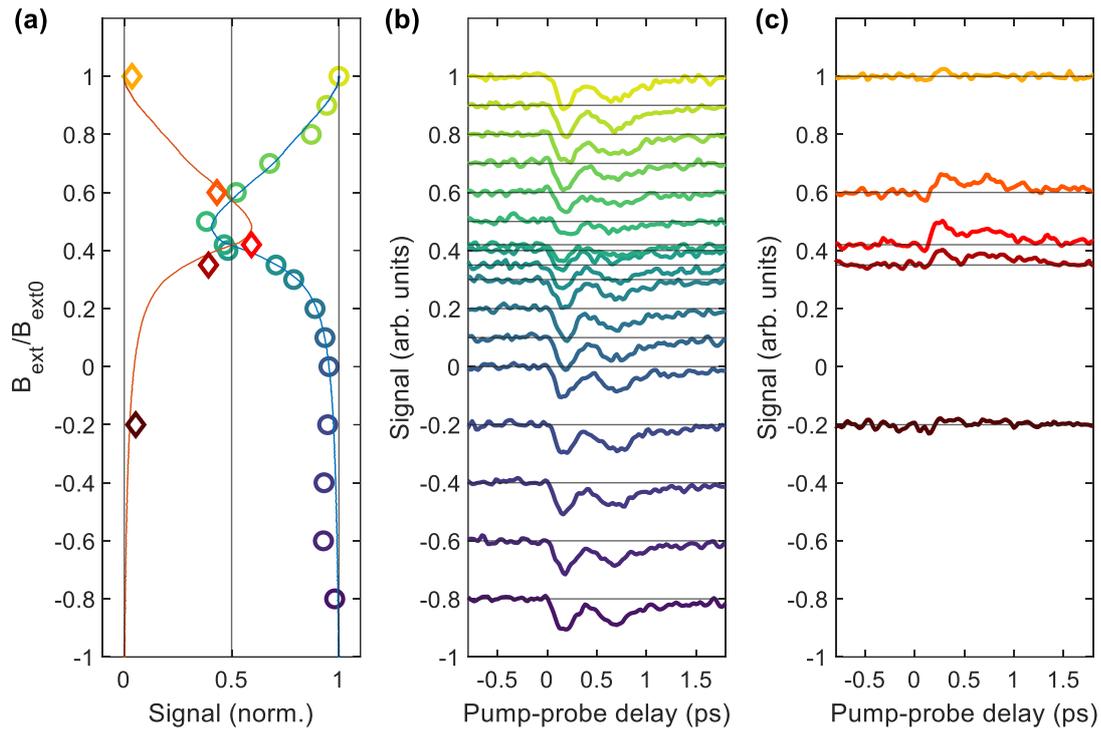

Figure S 7: Even-in-$B_{\mathrm{ext}}$ signals $\Delta S^+_{B_{\mathrm{ext}}}$ for $\varphi_{\mathrm{pr}} = 45°$ (basis for Fig. 5a). **(a)** Signal amplitudes for $\boldsymbol{E} \parallel \boldsymbol{u}_x$ (blue-green) and $\boldsymbol{E} \parallel \boldsymbol{u}_y$ (red-orange). Amplitudes were extracted according to Appendix B. **(b)** Signal traces $\Delta S(t)$ for $\boldsymbol{E} \parallel \boldsymbol{u}_x$ used to extract the amplitudes in panel (a). The top curve (light-green) shows the reference signal for the amplitude extraction. All signals are smoothed to remove high-frequency noise. Traces are shifted for clarity. (c) Signal traces $\Delta S(t)$ for $\boldsymbol{E} \parallel \boldsymbol{u}_y$ used to extract the amplitudes in panel (a).

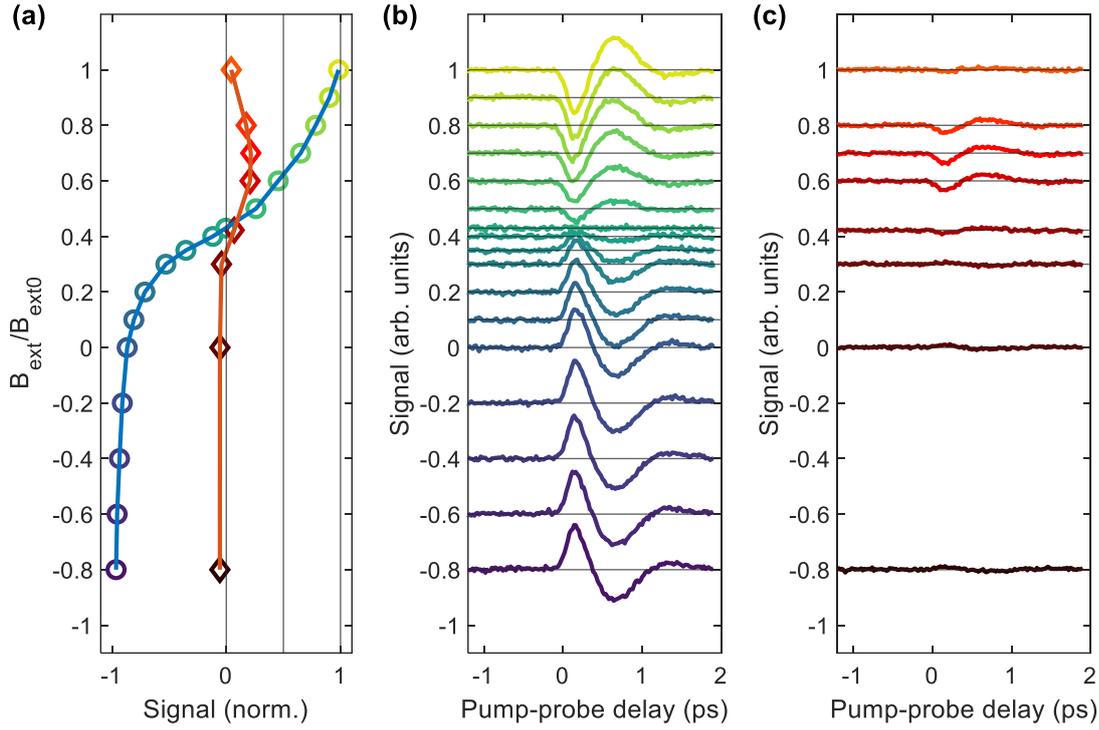

Figure S 8: Ellipticity signals $\left(\Delta S(\varphi_{\mathrm{pr}} = 0°) - \Delta S(\varphi_{\mathrm{pr}} = 90°)\right)/2$ odd in $\boldsymbol{B}_{\mathrm{ext}}$. **(a)** Signal amplitudes for $\boldsymbol{E} \parallel \boldsymbol{u}_x$ (blue-green) and $\boldsymbol{E} \parallel \boldsymbol{u}_y$ (red-orange). Amplitudes were extracted according to Appendix B. **(b)** Signal traces $\Delta S(t)$ for $\boldsymbol{E} \parallel \boldsymbol{u}_x$ used to extract the amplitudes in panel (a). The top curve (light-green) shows the reference signal for the amplitude extraction. All signals are smoothed to remove high-frequency noise. Traces are shifted for clarity. (c) Signal traces $\Delta S(t)$ for $\boldsymbol{E} \parallel \boldsymbol{u}_y$ used to extract the amplitudes in panel (a).

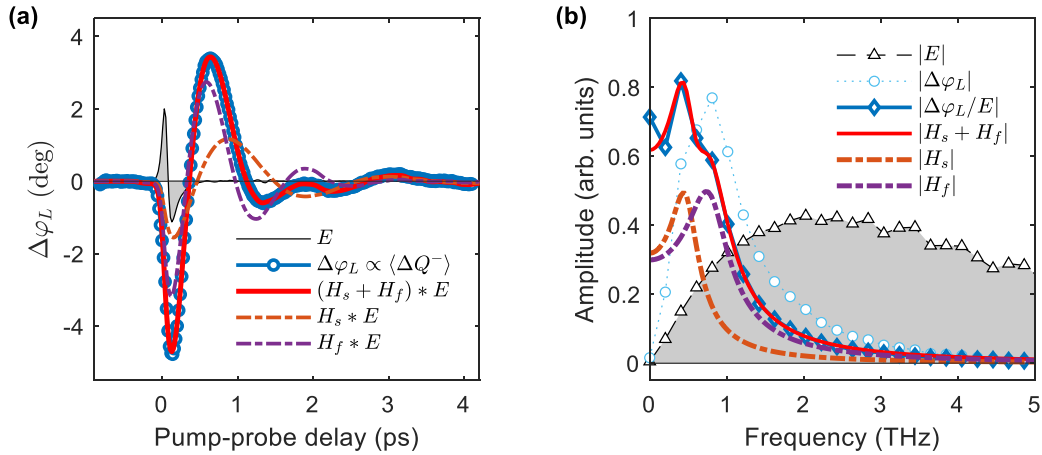

Figure S 9: Two-mode analysis. **(a)** Signal $\langle \Delta Q^- \rangle$ (Fig. 3a) with a fit based on $\Delta \varphi_L = (H_s + H_f) * E$ with a slow mode ($H_s$) and a fast mode ($H_f$) instead of a single mode. The black line with grey area shows $E(t)$ and blue circles the Néel-vector deflection angle $\Delta \varphi_L$. The red line shows the combined fit to $\Delta \varphi_L$, orange and purple dashed lines the individual mode contributions. **(b)** Amplitude of electric field (black triangles), experimental deflection angle (light-blue circles), experimentally obtained response (dark-blue diamonds), combined response (red line) and slow (orange, dashed) and fast mode (purple, dashed) response vs frequency.

Table 4: Mode parameters fitted with single effective mode and fast and slow mode. For comparison, the values extracted in ref. [4] for a bare film of Mn$_2$Au on Al$_2$O$_3$ are given. For the slow and fast mode, the fitted $A$ and, thus, the torkance $\lambda_{\mathrm{NSOT}}$ is an effective value for the assumption that the entire volume is uniformly excited and probed.

| | $H_{\mathrm{eff}}$ (Fig. 6) | $H_s$ (Figure S 9) | $H_f$ (Figure S 9) | $H_{\mathrm{eff}}$ (Ref. [4]) |
|---|---|---|---|---|
| **Input parameters** | | | | |
| $\sigma$ (MS m$^{-1}$) | 8.8 | 8.8 | 8.8 | 1.5 |
| **Fit parameters** | | | | |
| $\Omega_0/2\pi$ (THz) | 0.72 | 0.5 | 0.82 | 0.6 |
| $\Gamma$ (ps$^{-1}$) | 2.3 | 1.1 | 1.6 | 1.9 |
| $A$ (m V$^{-1}$ s$^{-1}$) | $2.1 \cdot 10^7$ | $0.5 \cdot 10^7$ | $1.3 \cdot 10^7$ | $0.5 \cdot 10^7$ |
| **Derived parameters** | | | | |
| $\alpha$ | $10 \cdot 10^{-2}$ | $5 \cdot 10^{-2}$ | $7 \cdot 10^{-2}$ | $8 \cdot 10^{-2}$ |
| $\lambda_{\mathrm{NSOT}}$ (cm$^2$ A$^{-1}$ s$^{-1}$) | 104 | 25* | 62* | 150 |

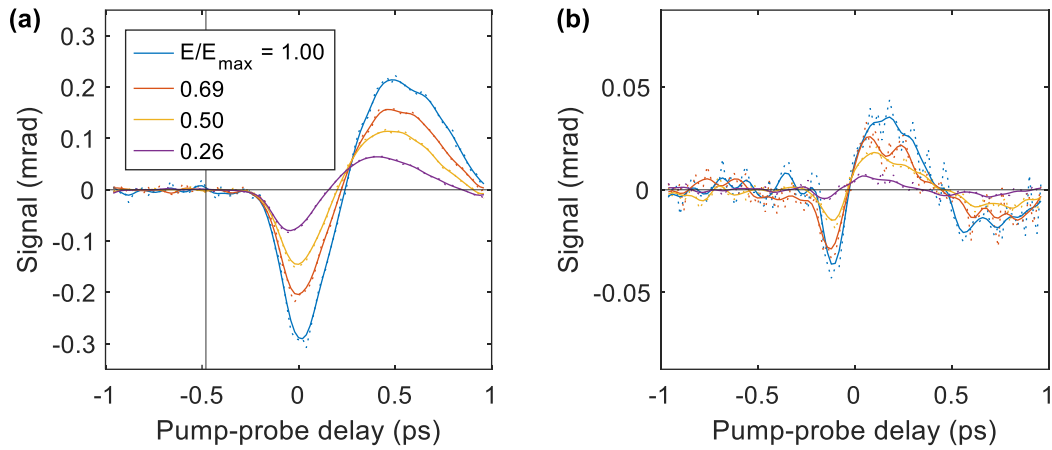

Figure S 10: Signal dependence of $\Delta S^-$ on the THz pump-field amplitude. **(a)** Signal components $\langle \Delta Q^- \rangle$. Dashed lines show the raw data, solid lines the low-pass filtered signal used for amplitude extraction. The noise amplitude in Fig 2b is the rms-amplitude of the low-pass-filtered signals before $-0.48$ ps (vertical line). **(b)** Signal components $\langle \Delta P^- \rangle$.

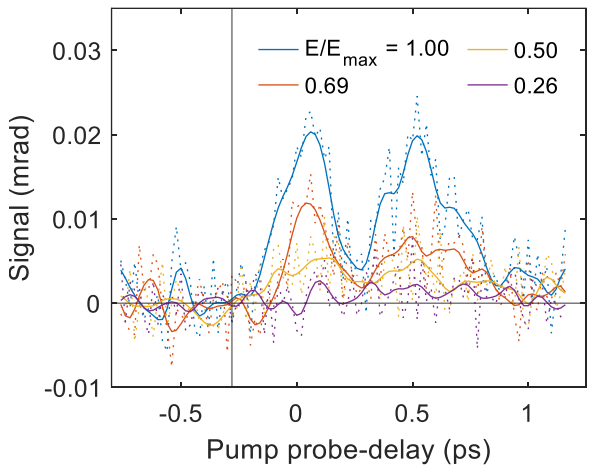

Figure S 11: Signal dependence of $\Delta \langle R^+ \rangle$ on THz pump field. Dashed lines show the raw data, solid lines the low-pass filtered signal used for amplitude extraction. The noise amplitude in Fig 3b is the rms-amplitude of the low-pass-filtered signals before $-0.28$ ps (vertical line).

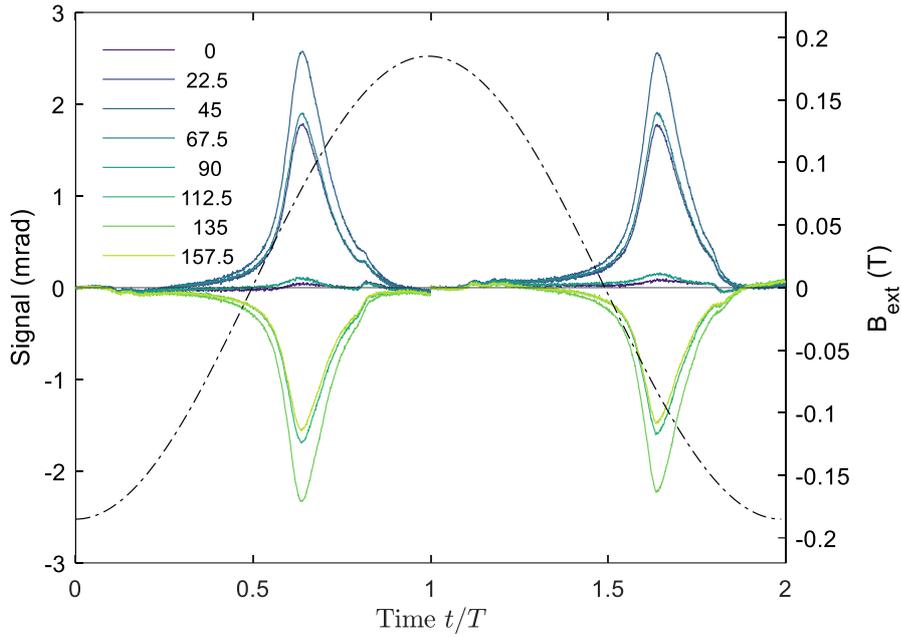

Figure S 12: Quasi-static probe rotation signals for 8 probe polarizations, where $\varphi_{\mathrm{pr}} = 0°$ corresponds to an $s$-polarized probe beam. The signal amplitude at $t = 0$, i.e. $B_{\mathrm{ext}} = -185$ mT, has been subtracted from each curve. Black dash-dotted line shows the external magnetic field $B_{\mathrm{ext}}(t)$.

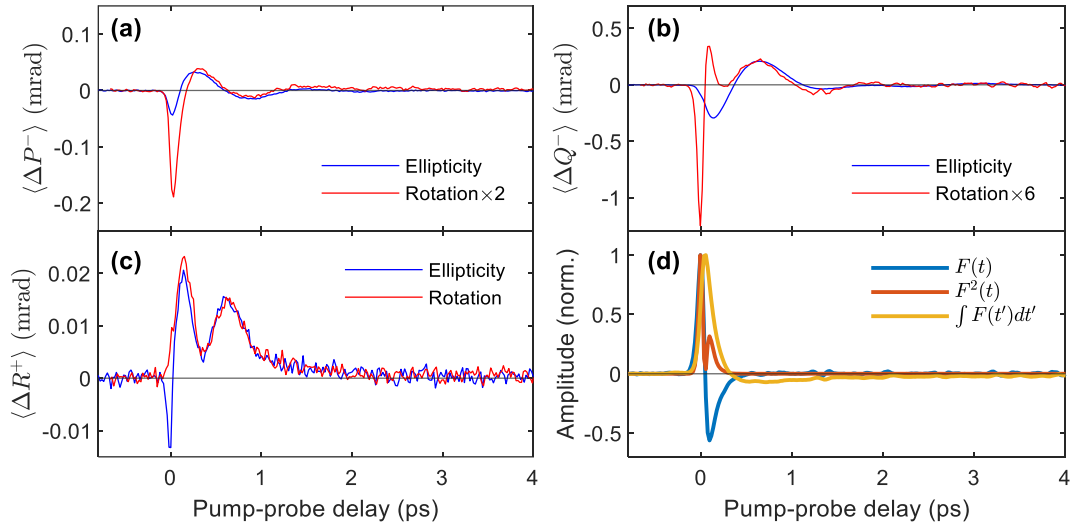

Figure S 13: Major signal components in rotation and ellipticity. **(a)** Contribution $\langle \Delta P^- \rangle$ for ellipticity (blue line) and rotation (red line, multiplied by 2). **(b)** Signal contribution $\langle \Delta Q^- \rangle$ for ellipticity (blue line) and rotation (red line, multiplied by 6). **(c)** Signal contribution $\langle \Delta R^+ \rangle$ for ellipticity (blue line) and rotation (red line). **(d)** Reference waveforms of the THz field $F(t)$, $F^2(t)$ and $\int_{-\infty}^{t} \mathrm{d}t' \, F(t')$.